\documentclass[12pt]{article}
\usepackage{amsmath,amssymb}
\usepackage{graphicx}
\newcommand{\eqdef}{\stackrel{\text{def}}{=}}
\newcommand{\n}{\nonumber\\}
\newcommand{\bm}{\boldsymbol}
\newcommand{\ignore}[1]{}
\renewcommand{\theequation}{\arabic{section}.\arabic{equation}}
\newcommand{\Romannumeral}[1]{\uppercase\expandafter{\romannumeral#1}}
\newcommand{\I}{\text{\Romannumeral{1}}}
\newcommand{\II}{\text{\Romannumeral{2}}}

\allowdisplaybreaks[4]

\begin{document}

\baselineskip=20pt

\newfont{\elevenmib}{cmmib10 scaled\magstep1}
\newcommand{\preprint}{
   \begin{flushleft}
     \elevenmib Yukawa\, Institute\, Kyoto\\
   \end{flushleft}\vspace{-1.3cm}
   \begin{flushright}\normalsize \sf
     DPSU-12-2\\
     YITP-12-48\\
   \end{flushright}}
\newcommand{\Title}[1]{{\baselineskip=26pt
   \begin{center} \Large \bf #1 \\ \ \\ \end{center}}}
\newcommand{\Author}{\begin{center}
   \large \bf Satoru Odake${}^a$ and Ryu Sasaki${}^b$ \end{center}}
\newcommand{\Address}{\begin{center}
     $^a$ Department of Physics, Shinshu University,\\
     Matsumoto 390-8621, Japan\\
     ${}^b$ Yukawa Institute for Theoretical Physics,\\
     Kyoto University, Kyoto 606-8502, Japan
   \end{center}}
\newcommand{\Accepted}[1]{\begin{center}
   {\large \sf #1}\\ \vspace{1mm}{\small \sf Accepted for Publication}
   \end{center}}

\preprint
\thispagestyle{empty}

\Title{Multi-indexed Wilson and Askey-Wilson Polynomials}

\Author

\Address
\vspace{1cm}

\begin{abstract}
As the third stage of the project {\em multi-indexed orthogonal
polynomials\/}, we present, in the framework of `discrete quantum
mechanics' with pure imaginary shifts in one dimension, the multi-indexed
Wilson and Askey-Wilson polynomials. They are obtained from the
original Wilson and Askey-Wilson polynomials by multiple application
of the discrete analogue of the Darboux transformations or the
Crum-Krein-Adler deletion of `virtual state solutions' of type $\I$
and $\II$, in a similar way to the multi-indexed Laguerre, Jacobi
and ($q$-)Racah polynomials reported earlier.
\end{abstract}

\section{Introduction}
\label{sec:intro}

This is a third report of the project {\em multi-indexed orthogonal
polynomials\/}. Following the examples of multi-indexed Laguerre and
Jacobi polynomials \cite{os25}, multi-indexed ($q$-)Racah polynomials
\cite{os26}, we present multi-indexed Wilson and Askey-Wilson polynomials
constructed in the framework of discrete quantum mechanics with pure
imaginary shifts \cite{os13}. It is well-known  that the original Wilson
and Askey-Wilson polynomials are the most generic members of the Askey
scheme of hypergeometric orthogonal polynomials
\cite{askey,ismail,koeswart,gasper}.
These new multi-indexed orthogonal polynomials are specified by a set of
indices $\mathcal{D}=\{d_1,\ldots,d_M\}$ consisting of distinct natural
numbers $d_j\in\mathbb{N}$, on top of $n$, which counts the nodes as in
the ordinary orthogonal polynomials. The simplest examples,
$\mathcal{D}=\{\ell\}$, $\ell\ge1$, $\{P_{\ell,n}(x)\}$ are also called
{\em exceptional orthogonal polynomials\/} \cite{gomez}--\cite{quesne3}.
They are obtained as the main part of the eigenfunctions (vectors) of
various {\em exactly solvable\/} Schr\"odinger equations in one dimensional
quantum mechanics and their `discrete' generalisations, in which the
corresponding Schr\"odinger equations are second order difference equations
\cite{os13,os12,os14}. They form a complete set of orthogonal polynomials,
although they start at a certain positive degree ($\ell\ge1$) rather than
a degree zero constant term. The latter situation is essential for avoiding
the constraints of Bochner's theorem \cite{bochner}.
We strongly believe that these new orthogonal polynomials will find
plenty of novel applications in various branches of science and technology
in the good old tradition of orthogonal polynomials.

The basic logic for constructing multi-indexed orthogonal polynomials
is essentially the same for the ordinary Schr\"odinger equations,
{\em i.e.\/} those for the Laguerre and Jacobi polynomials and for
the difference Schr\"odinger equations with real as well as pure imaginary
shifts, {\em i.e.\/} the ($q$-)Racah polynomials and the Wilson and
Askey-Wilson polynomials, etc.
The main ingredients are the factorised Hamiltonians, the Crum-Krein-Adler
formulas \cite{crum, adler, Nsusy} for deletion of eigenstates,
{\em that is\/} the multiple Darboux transformations \cite{darb} and
the {\em virtual state solutions} \cite{os25} which are generated by
twisting the discrete symmetries of the original Hamiltonians.
Most of these methods for discrete Schr\"odinger equations had been
developed \cite{os12,os24,os13,os14,os15,gos,os22} and they were used
for the exceptional Wilson and Askey-Wilson polynomials \cite{os17,os20}.
In different contexts, Darboux transformations for orthogonal polynomials
have been discussed by many authors \cite{grun-haine,grun-yakim,haine-ill}.

It is important to stress that the factorised Hamiltonians in the
discrete quantum mechanics, that is, those governing the ($q$-)Racah,
Wilson, Askey-Wilson polynomials etc possess certain discrete symmetry.
They lead to virtual Hamiltonians which are linearly connected with
the original Hamiltonian. (See \eqref{propV'}--\eqref{H'} of the present
paper and (2.18)--(2.22), (2.59)--(2.63) of \cite{os26}.)
In the ordinary quantum mechanics of the radial oscillator potential
($x^2+\frac{g(g-1)}{x^2}$, for the Laguerre polynomials) and the
P\"oschl-Teller potential ($\frac{g(g-1)}{\sin^2x}+\frac{h(h-1)}{\cos^2x}$,
for the Jacobi polynomials), the discrete symmetry is well known.
For the former, $g\to 1-g$ and/or $x\to ix$ and for the latter $g\to 1-g$
and/or $h\to 1-h$. 
To the best of our knowledge, those discrete symmetries for the
($q$-)Racah, Wilson and Askey-Wilson systems do not seem to be widely
recognised, since they are not easily identifiable in the polynomial
equations. 
The virtual state solutions belong to the virtual Hamiltonians.
The actual contents of virtual state solutions depend on the types of
the Schr\"odinger equations. For the ordinary Schr\"odinger equations
with a second order differential operator, the virtual state solutions
satisfy the Schr\"odinger equation. But they do not belong to the Hilbert
space of square integrable solutions, due to the twisted boundary condition
on either one of the two boundaries, to be called the type $\I$ or $\II$.
For the multi-indexed ($q$-)Racah polynomials in the discrete quantum
mechanics with real shifts, the virtual state `solutions' fail to satisfy
the Schr\"odinger equation at either one of the two boundary points
\cite{os26}. They are called virtual state vectors of type $\I$ or $\II$.
In the present case of discrete quantum mechanics with pure imaginary shifts,
the virtual state solutions satisfy the difference Schr\"odinger equation.
But they do not belong to the Hilbert space of eigenfunctions either by
the lack of square integrability or by the presence of singularities in
certain rectangular domain. (See more detailed discussion in section
\ref{sec:analy}.)
In other words, the analyticity requirements supersede the boundary
conditions which used to classify the virtual state solutions for the
Laguerre, Jacobi and ($q$-)Racah cases.
In section three, we will introduce two types of twistings or the discrete
symmetry transformations and the corresponding virtual Hamiltonians and
virtual state solutions. They are of the same structure but adopting
different sets of parameters. We will call them of type $\I$ and $\II$
as in the other cases but they are not related to boundary conditions.
In all these cases, the features disqualifying them to become the
eigenfunctions are carried by the so called ``virtual groundstate''
functions $\tilde{\phi}_0(x)$, \eqref{xiv=}.
The polynomial part of the virtual state solutions, to be denoted by
$\xi_\text{v}(\eta)$ \eqref{xiv=}, are the genuine solutions of equations
determining the eigenpolynomials, but with twisted parameters.
It is the virtual state polynomials $\{\xi_\text{v}(\eta)\}$, 
not the virtual groundstate $\tilde{\phi}_0(x)$, that play the main role
in the construction of
multi-indexed and exceptional \cite{os17,os20} polynomials and the set
of their degrees $\{d_1,\ldots,d_M\}$ constitutes the {\em multi-index\/}.
We focus on the algebraic structure of the multi-indexed
orthogonal polynomials and their difference equations, which hold for
any parameter range.
We do not pursue the other important aspect of the problem,
that is the determination of the parameter ranges in which the hermiticity
of the multi-indexed Hamiltonians and the positivity of the orthogonality
weight functions for the multi-indexed polynomials are ensured.

This paper is organised as follows.
In section two, the basic logic of virtual states deletion in discrete
quantum mechanics with pure imaginary shifts in general is outlined.
Starting from the general setting of discrete quantum mechanics with
pure imaginary shifts in \S\,\ref{sec:ori}, the analyticity requirements
in connection with the hermiticity (self-adjointness) of the Hamiltonians
are briefly recapitulated in \S\,\ref{sec:analy}.
General procedures and formulas of multiple virtual states deletion are
reviewed in \S\,\ref{sec:del_virtual}. The main logics are essentially
the same as those for the multi-indexed Laguerre, Jacobi and ($q$-)Racah
polynomials but the explicit formulas look rather different reflecting
the specific properties of discrete quantum mechanics with pure imaginary
shifts.
After recapitulating the basic properties of the Wilson and Askey-Wilson
systems in \S\,\ref{sec:org_AW}, the discrete symmetries of the the
Wilson and Askey-Wilson systems are introduced in \S\,\ref{sec:discr}.
The multi-indexed Wilson and Askey-Wilson polynomials are constructed
explicitly in \S\,\ref{sec:miop_AW} for the type $\I$ and $\II$ virtual
states deletions.
The analyticity and hermiticity of the multi-indexed hamiltonians are
discussed in some detail in section \S\ref{sec:analherm}.
The final section is for a summary and comments including the limits to
the multi-indexed Jacobi and Laguerre polynomials.
For simplicity of presentation we relegate several technical results
to Appendix.

Throughout this paper we will focus on the algebraic aspects of the theory.
To determine the exact ranges of validity of various formulas is another
problem.

\section{Formulation}
\label{sec:form}
\setcounter{equation}{0}

\subsection{Original system}
\label{sec:ori}

Let us recapitulate the discrete quantum mechanics with pure imaginary
shifts developed in \cite{os13}.
The dynamical variables are the real coordinate $x$ ($x_1\leq x\leq x_2$) and
the conjugate momentum $p=-i\partial_x$, which are governed by the
following factorised positive semi-definite Hamiltonian:
\begin{align}
  &\mathcal{H}\eqdef\sqrt{V(x)}\,e^{\gamma p}\sqrt{V^*(x)}
  +\!\sqrt{V^*(x)}\,e^{-\gamma p}\sqrt{V(x)}
  -V(x)-V^*(x)=\mathcal{A}^{\dagger}\mathcal{A},
  \label{H}\\
  &\mathcal{A}\eqdef i\bigl(e^{\frac{\gamma}{2}p}\sqrt{V^*(x)}
  -e^{-\frac{\gamma}{2}p}\sqrt{V(x)}\,\bigr),\quad
  \mathcal{A}^{\dagger}\eqdef -i\bigl(\sqrt{V(x)}\,e^{\frac{\gamma}{2}p}
  -\sqrt{V^*(x)}\,e^{-\frac{\gamma}{2}p}\bigr).
\end{align}
Here the potential function $V(x)$ is an analytic function of $x$ and
$\gamma$ is a real constant.
The $*$-operation on an analytic function $f(x)=\sum_na_nx^n$
($a_n\in\mathbb{C}$) is defined by $f^*(x)=\sum_na_n^*x^n$, in which
$a_n^*$ is the complex conjugation of $a_n$.
Obviously $f^{**}(x)=f(x)$ and $f(x)^*=f^*(x^*)$.
If a function satisfies $f^*=f$, then it takes real values on the real line.
Since the momentum operator appears in exponentiated forms,
the Schr\"{o}dinger equation
\begin{equation}
  \mathcal{H}\phi_n(x)=\mathcal{E}_n\phi_n(x)
  \ \ (n=0,1,2,\ldots),
\end{equation}
is an analytic difference equation with pure imaginary shifts instead
of a differential equation.
Throughout this paper we consider those systems which have a
square-integrable groundstate together with an infinite number of
discrete energy levels:
$0=\mathcal{E}_0 <\mathcal{E}_1 < \mathcal{E}_2 < \cdots$.
The orthogonality relation reads
\begin{equation}
  (\phi_n,\phi_m)\eqdef
  \int_{x_1}^{x_2}\!\!dx\,\phi_n^*(x)\phi_m(x)=h_n\delta_{nm}
  \ \ (n,m=0,1,2,\ldots),\quad 0<h_n<\infty.
\end{equation}
The eigenfunctions $\phi_n(x)$ can be chosen `real', $\phi_n^*(x)=\phi_n(x)$,
and the groundstate wavefunction $\phi_0(x)$ is determined as the zero
mode of the operator $\mathcal{A}$, $\mathcal{A}\phi_0(x)=0$, namely,
\begin{equation}
  \sqrt{V^*(x-i\tfrac{\gamma}{2})}\,\phi_0(x-i\tfrac{\gamma}{2})
  =\sqrt{V(x+i\tfrac{\gamma}{2})}\,\phi_0(x+i\tfrac{\gamma}{2}).
  \label{Aphi0=0}
\end{equation}

\subsection{Analyticity requirements}
\label{sec:analy}

The hermiticity of the Hamiltonians of discrete quantum mechanics with
pure imaginary shifts is more involved than that of the ordinary quantum
mechanics \cite{os13,os14}.
Here we review the hermiticity of the Hamiltonian \eqref{H} in general,
in a way applicable to those appearing in the multi-indexed Wilson and
Askey-Wilson systems, {\em e.g.} \eqref{Hd1..ds}--\eqref{Vhd1..ds},
\eqref{sstanham1}--\eqref{Vd1..ds}.
Of course, the hermiticity of the original Hamiltonians of the Wilson and
Askey-Wilson systems \eqref{Vform}--\eqref{hn} is well established
\cite{os13,os14}.

Let us consider the functions of the form
$f(x)=\phi_0(x)\check{\mathcal{R}}(x)$, where
$\phi_0(x)^2$ and $\check{\mathcal{R}}(x)$ are meromorphic functions and
$\check{\mathcal{R}}^*(x)=\check{\mathcal{R}}(x)$.
For two such functions $f_1=\phi_0\check{\mathcal{R}}_1$ and
$f_2=\phi_0\check{\mathcal{R}}_2$, the condition of the hermiticity
$(f_1,\mathcal{H}f_2)=(\mathcal{H}f_1,f_2)$ becomes
\begin{equation}
  \int_{x_1}^{x_2}dx\bigl(G(x-i\tfrac{\gamma}{2})
  +G^*(x+i\tfrac{\gamma}{2})\bigr)
  =\int_{x_1}^{x_2}dx\bigl(G(x+i\tfrac{\gamma}{2})
  +G^*(x-i\tfrac{\gamma}{2})\bigr),
  \label{cond(iii)}
\end{equation}
where $G(x)$ is defined by
\begin{align}
  &G(x)=V(x+i\tfrac{\gamma}{2})\phi_0(x+i\tfrac{\gamma}{2})^2
  \check{\mathcal{R}}_1(x+i\tfrac{\gamma}{2})
  \check{\mathcal{R}}_2(x-i\tfrac{\gamma}{2}),\n
  &\Bigl(\Rightarrow
  G^*(x)=V(x+i\tfrac{\gamma}{2})\phi_0(x+i\tfrac{\gamma}{2})^2
  \check{\mathcal{R}}_1(x-i\tfrac{\gamma}{2})
  \check{\mathcal{R}}_2(x+i\tfrac{\gamma}{2})\Bigr).
  \label{Gdef}
\end{align}
Although the term $V(x)+V^*(x)$ in $\mathcal{H}$ are canceled out in
this calculation, this term $V(x)+V^*(x)$ should be non-singular for
$x_1\leq x\leq x_2$, and the integral $\int_{x_1}^{x_2}dx\,V(x)\phi_0(x)^2
\check{\mathcal{R}}_1(x)\check{\mathcal{R}}_2(x)$ should be finite.
By using the residue theorem, the condition \eqref{cond(iii)} is rewritten as
\begin{align}
  &\quad\int_{-\frac{\gamma}{2}}^{\frac{\gamma}{2}}dx
  \bigl(G(x_2+ix)-G^*(x_2-ix)\bigr)
  -\int_{-\frac{\gamma}{2}}^{\frac{\gamma}{2}}dx
  \bigl(G(x_1+ix)-G^*(x_1-ix)\bigr)\n
  &=2\pi\frac{\gamma}{|\gamma|}
  \sum_{x_0:\text{pole in $D_{\gamma}$}}\!\!\!\text{Res}_{x_0}
  \bigl(G(x)-G^*(x)\bigr),
  \label{cond(iii)'}
\end{align}
where the residue of the function $G(x)-G^*(x)$ is taken at the poles
in the rectangular domain $D_{\gamma}$ :
\begin{equation}
  D_{\gamma}\eqdef\bigl\{x\in\mathbb{C}\bigm|x_1\leq\text{Re}\,x\leq x_2,
  |\text{Im}\,x|\leq\tfrac12|\gamma|\bigr\}.
  \label{Dgamma}
\end{equation}

In our previous work on the exceptional Wilson and  Askey-Wilson
\cite{os17,os20}, we required that $G$ and $G^*$ have no poles in the
rectangular domain $D_{\gamma}$, as a sufficient condition for the
hermiticity of the Hamiltonian.
This was too strong a requirement.
In later examples,
$\check{\mathcal{R}}(x)=\mathcal{R}\bigl(\eta(x)\bigr)$
is a rational function of $\eta(x)$ and $V(x)\phi_0(x)^2$ has the form
$\sim (V\phi_0^2)_\text{original}\times(\text{rational function of
$\eta(x)$})$.
In the original Wilson and Askey Wilson theory, $(V\phi_0^2)_\text{original}$
part has no poles in $D_{\gamma}$, \eqref{Vphi0^2}.
In the deformed theory in general, however,
$(V\phi_0^2)_\text{original}$ has shifted parameters and we continue to
require that this part has no poles in the rectangular domain $D_{\gamma}$.
We also remark that even if the (rational function of $\eta(x)$)-part has
poles in the rectangular domain $D_{\gamma}$, there is a possibility that
the sum of the residues vanish.

\subsection{Deletion of virtual states}
\label{sec:del_virtual}

In \cite{gos} we have presented the Crum-Adler scheme, {\em i.e.\/}
the deletion of $M$ eigenstates. In that case the index set of the
deleted eigenstates $\mathcal{D}\eqdef\{d_1,d_2,\ldots,d_M\}$
($d_j\in\mathbb{Z}_{\geq 0}$) should satisfy the condition
$\prod_{j=1}^M(m-d_j)\ge0$ ($\forall m\in\mathbb{Z}_{\ge 0}$), eq.\,(2.8)
in \cite{gos}.
We now apply the Crum-Adler scheme to virtual states instead of eigenstates.
The above condition eq.\,(2.8) in \cite{gos} is no longer necessary.

The Casorati determinant of a set of $n$ functions $\{f_j(x)\}$ is defined
by
\begin{equation}
  \text{W}_{\gamma}[f_1,\ldots,f_n](x)
  \eqdef i^{\frac12n(n-1)}
  \det\Bigl(f_k\bigl(x^{(n)}_j\bigr)\Bigr)_{1\leq j,k\leq n},\quad
  x_j^{(n)}\eqdef x+i(\tfrac{n+1}{2}-j)\gamma,
  \label{Wdef}
\end{equation}
\ignore{
\begin{equation}
  \lim_{\gamma\to 0}\gamma^{-\frac12n(n-1)}
  \text{W}_{\gamma}[f_1,f_2,\ldots,f_n](x)
  =\text{W}[f_1,f_2,\ldots,f_n](x),
\end{equation}
}
(for $n=0$, we set $\text{W}_{\gamma}[\cdot](x)=1$),
which satisfies identities
\begin{align}
  &\text{W}_{\gamma}[f_1,\ldots,f_n]^*(x)
  =\text{W}_{\gamma}[f_1^*,\ldots,f_n^*](x),\\
  &\text{W}_{\gamma}[gf_1,gf_2,\ldots,gf_n]
  =\prod_{j=1}^ng\bigl(x^{(n)}_j\bigr)\cdot
  \text{W}_{\gamma}[f_1,f_2,\ldots,f_n](x),
  \label{dWformula1}\\
  &\text{W}_{\gamma}\bigl[\text{W}_{\gamma}[f_1,f_2,\ldots,f_n,g],
  \text{W}_{\gamma}[f_1,f_2,\ldots,f_n,h]\,\bigr](x)\n
  &=\text{W}_{\gamma}[f_1,f_2,\ldots,f_n](x)\,
  \text{W}_{\gamma}[f_1,f_2,\ldots,f_n,g,h](x)
  \quad(n\geq 0).
  \label{dWformula2}
\end{align}

Let us assume the existence of an analytic function $V'(x)$ of $x$
satisfying
\begin{align}
  &V(x)V^*(x-i\gamma)=\alpha^2V'(x)V^{\prime*}(x-i\gamma),
  \qquad\quad\alpha>0,\n
  &V(x)+V^*(x)=\alpha\bigl(V'(x)+V^{\prime*}(x)\bigr)-\alpha',
  \qquad\alpha'<0,
  \label{propV'}
\end{align}
where $\alpha$ and $\alpha'$ are constants.
Then we obtain a linear relation between two Hamiltonians:
\begin{align}
  \mathcal{H}&=\alpha\mathcal{H}'+\alpha',
  \label{H=aH'+a'}\\
  \mathcal{H}'&\eqdef\sqrt{V'(x)}\,e^{\gamma p}\sqrt{V^{\prime*}(x)}
  +\!\sqrt{V^{\prime*}(x)}\,e^{-\gamma p}\sqrt{V'(x)}
  -V'(x)-V^{\prime*}(x).
  \label{H'}
\end{align}
Since $\mathcal{H}$ is positive semi-definite, $\mathcal{H}'$ is
obviously positive definite and it has no zero-mode.
Let us also assume the existence of {\em virtual state wavefunctions}
$\tilde{\phi}_{\text{v}}(x)$ ($\text{v}\in\mathcal{V}$), which are
`polynomial solutions' of degree $\text{v}$ of the Schr\"odinger equation
\begin{equation}
  \mathcal{H}\tilde{\phi}_{\text{v}}(x)
  =\tilde{\mathcal{E}}_{\text{v}}\tilde{\phi}_{\text{v}}(x)
  \ \ \text{or}\ \ \mathcal{H}'\tilde{\phi}_{\text{v}}(x)
  =\mathcal{E}'_{\text{v}}\tilde{\phi}_{\text{v}}(x),\quad
  \tilde{\mathcal{E}}_{\text{v}}\eqdef\alpha\mathcal{E}'_{\text{v}}+\alpha',
  \quad
  \tilde{\phi}^*_{\text{v}}(x)=\tilde{\phi}_{\text{v}}(x),
  \label{Hphiv=}
\end{equation}
but they, including the zeromode $\tilde{\phi}_0$, do not belong to
the Hilbert space of $\mathcal{H}$.
Here $\mathcal{V}$ is the index set of the virtual state wavefunctions.
We require that $\tilde{\mathcal{E}}_{\text{v}}<0$ and
some analytic properties of $\tilde{\phi}_{\text{v}}(x)$,
which are explicitly presented in \S\,\ref{sec:miop}.

We have developed the method of virtual states deletion for ordinary
quantum mechanics in \cite{os25} and for discrete quantum mechanics
with real shifts in \cite{os26}.
Algebraic aspects of this method are the same and can be applied to discrete
quantum mechanics with pure imaginary shifts.
The procedure is as follows;
(\romannumeral1) rewrite the original Hamiltonian as
$\mathcal{H}=\hat{\mathcal{A}}_{d_1}^{\dagger}\hat{\mathcal{A}}_{d_1}
+\tilde{\mathcal{E}}_{d_1}$ ($d_1\in\mathcal{V}$),
(\romannumeral2) define a new isospectral Hamiltonian
$\mathcal{H}_{d_1}\eqdef\hat{\mathcal{A}}_{d_1}
\hat{\mathcal{A}}_{d_1}^{\dagger}+\tilde{\mathcal{E}}_{d_1}$,
whose eigenfunctions are given by
$\phi_{d_1n}(x)\eqdef\hat{\mathcal{A}}_{d_1}\phi_n(x)$
together with virtual state wavefunctions
$\tilde{\phi}_{d_1\text{v}}(x)\eqdef\hat{\mathcal{A}}_{d_1}
\tilde{\phi}_{\text{v}}(x)$ ($\text{v}\in\mathcal{V}\backslash\{d_1\}$),
$\mathcal{H}_{d_1}\phi_{d_1n}(x)=\mathcal{E}_n\phi_{d_1n}(x)$,
$\mathcal{H}_{d_1}\tilde{\phi}_{d_1\text{v}}(x)
=\tilde{\mathcal{E}}_\text{v}\tilde{\phi}_{d_1\text{v}}(x)$
(\romannumeral3) rewrite this as
$\mathcal{H}_{d_1}=\hat{\mathcal{A}}_{d_1d_2}^{\dagger}
\hat{\mathcal{A}}_{d_1d_2}+\tilde{\mathcal{E}}_{d_2}$
($d_2\in\mathcal{V}\backslash\{d_1\}$),
(\romannumeral4) define the next isospectral Hamiltonian
$\mathcal{H}_{d_1d_2}\eqdef\hat{\mathcal{A}}_{d_1d_2}
\hat{\mathcal{A}}_{d_1d_2}^{\dagger}+\tilde{\mathcal{E}}_{d_2}$,
whose eigenfunctions are given by
$\phi_{d_1d_2n}(x)\eqdef\hat{\mathcal{A}}_{d_1d_2}\phi_{d_1n}(x)$
together with virtual state wavefunctions
$\tilde{\phi}_{d_1d_2\text{v}}(x)\eqdef\hat{\mathcal{A}}_{d_1d_2}
\tilde{\phi}_{d_1\text{v}}(x)$
($\text{v}\in\mathcal{V}\backslash\{d_1,d_2\}$),
$\mathcal{H}_{d_1d_2}\phi_{d_1d_2n}(x)=\mathcal{E}_n\phi_{d_1d_2n}(x)$,
$\mathcal{H}_{d_1d_2}\tilde{\phi}_{d_1d_2\text{v}}(x)
=\tilde{\mathcal{E}}_\text{v}\tilde{\phi}_{d_1d_2\text{v}}(x)$,
(\romannumeral5) by repeating this process, we obtain
$\mathcal{H}_{d_1\ldots d_s}$ and
its eigenfunctions $\phi_{d_1\ldots d_s\,n}(x)$ together with virtual
state wavefunctions $\tilde{\phi}_{d_1\ldots d_s\,\text{v}}(x)$,
(\romannumeral6) $\mathcal{H}_{d_1\ldots d_s}$ can be written in the standard
form, $\mathcal{H}_{d_1\ldots d_s}=\mathcal{A}_{d_1\ldots d_s}^{\dagger}
\mathcal{A}_{d_1\ldots d_s}$.
If the resulting system is well-defined, we obtain the isospectrally
deformed systems just  as those in refs.\cite{os25} and \cite{os26}.

Here we present `formal' expressions of the deformed systems, which are
proved inductively.
The system obtained after $s$ virtual state deletions ($s\geq 1$),
which are labeled by $\{d_1,\ldots,d_s\}$
($d_j\in\mathcal{V}$ : mutually distinct), is
\begin{align}
  &\mathcal{H}_{d_1\ldots d_s}\eqdef
  \hat{\mathcal{A}}_{d_1\ldots d_s}\hat{\mathcal{A}}_{d_1\ldots d_s}^{\dagger}
  +\tilde{\mathcal{E}}_{d_s},
  \label{Hd1..ds}\\
  &\hat{\mathcal{A}}_{d_1\ldots d_s}\eqdef
  i\bigl(e^{\frac{\gamma}{2}p}\sqrt{\hat{V}_{d_1\ldots d_s}^*(x)}
  -e^{-\frac{\gamma}{2}p}\sqrt{\hat{V}_{d_1\ldots d_s}(x)}\,\bigr),\n
  &\hat{\mathcal{A}}_{d_1\ldots d_s}^{\dagger}\eqdef
  -i\bigl(\sqrt{\hat{V}_{d_1\ldots d_s}(x)}\,e^{\frac{\gamma}{2}p}
  -\sqrt{\hat{V}_{d_1\ldots d_s}^*(x)}\,e^{-\frac{\gamma}{2}p}\bigr),\\
  &\hat{V}_{d_1\ldots d_s}(x)\eqdef
  \sqrt{V(x-i\tfrac{s-1}{2}\gamma)V^*(x-i\tfrac{s+1}{2}\gamma)}\n
  &\phantom{\hat{V}_{d_1\ldots d_s}(x)\eqdef}\times
  \frac{\text{W}_{\gamma}[\tilde{\phi}_{d_1},\ldots,\tilde{\phi}_{d_{s-1}}]
  (x+i\frac{\gamma}{2})}
  {\text{W}_{\gamma}[\tilde{\phi}_{d_1},\ldots,\tilde{\phi}_{d_{s-1}}]
  (x-i\frac{\gamma}{2})}\,
  \frac{\text{W}_{\gamma}[\tilde{\phi}_{d_1},\ldots,\tilde{\phi}_{d_s}]
  (x-i\gamma)}
  {\text{W}_{\gamma}[\tilde{\phi}_{d_1},\ldots,\tilde{\phi}_{d_s}](x)},
  \label{Vhd1..ds}\\
  &\phi_{d_1\ldots d_s\,n}(x)\eqdef
  \hat{\mathcal{A}}_{d_1\ldots d_s}\phi_{d_1\ldots d_{s-1}\,n}(x)
  \ \ (n=0,1,2,\ldots),\n
  &\tilde{\phi}_{d_1\ldots d_s\,\text{v}}(x)\eqdef
  \hat{\mathcal{A}}_{d_1\ldots d_s}
  \tilde{\phi}_{d_1\ldots d_{s-1}\,\text{v}}(x)
  \ (\text{v}\in\mathcal{V}\backslash\{d_1,\ldots,d_s\}),\\
  &\mathcal{H}_{d_1\ldots d_s}\phi_{d_1\ldots d_s\,n}(x)
  =\mathcal{E}_n\phi_{d_1\ldots d_s\,n}(x)
  \ \ (n=0,1,2,\ldots),\n
  &\mathcal{H}_{d_1\ldots d_s}\tilde{\phi}_{d_1\ldots d_s\,\text{v}}(x)
  =\tilde{\mathcal{E}}_\text{v}\tilde{\phi}_{d_1\ldots d_s\,\text{v}}(x)
  \ \ (\text{v}\in\mathcal{V}\backslash\{d_1,\ldots,d_s\}),
  \label{Hd1..dsphid1..ds=}\\
  &(\phi_{d_1\ldots d_s\,n},\phi_{d_1\ldots d_s\,m})
  =\prod_{j=1}^s(\mathcal{E}_n-\tilde{\mathcal{E}}_{d_j})\cdot
  h_n\delta_{nm}
  \ \ (n,m=0,1,2,\ldots).
  \label{(phid1..dsm,phid1..dsn)}
\end{align}
Let us remark that the eigenfunctions and the virtual state solutions
in all steps are `real' by construction,
$\phi_{d_1\ldots d_s\,n}^*(x)=\phi_{d_1\ldots d_s\,n}(x)$,
$\tilde{\phi}_{d_1\ldots d_s\,\text{v}}^*(x)
=\tilde{\phi}_{d_1\ldots d_s\,\text{v}}(x)$
and they have Casoratian expressions:
\begin{align}
  &\phi_{d_1\ldots d_s\,n}(x)=A(x)
  \text{W}_{\gamma}[\tilde{\phi}_{d_1},\ldots,\tilde{\phi}_{d_s},\phi_n](x),
  \n
  &\tilde{\phi}_{d_1\ldots d_s\,\text{v}}(x)=A(x)
  \text{W}_{\gamma}[\tilde{\phi}_{d_1},\ldots,\tilde{\phi}_{d_s},
  \tilde{\phi}_{\text{v}}](x),
  \label{phid1..dsn}\\
  &\quad A(x)=\left(
  \frac{\sqrt{\prod_{j=0}^{s-1}V(x+i(\frac{s}{2}-j)\gamma)
  V^*(x-i(\frac{s}{2}-j)\gamma)}}
  {\text{W}_{\gamma}[\tilde{\phi}_{d_1},\ldots,\tilde{\phi}_{d_s}]
  (x-i\frac{\gamma}{2})
  \text{W}_{\gamma}[\tilde{\phi}_{d_1},\ldots,\tilde{\phi}_{d_s}]
  (x+i\frac{\gamma}{2})}\right)^{\frac12},
  \nonumber
\end{align}
which are shown by using \eqref{dWformula2}.

Writing down \eqref{Hd1..dsphid1..ds=} and dividing it by
$\phi_{d_1\ldots d_s\,n}(x)$ or $\tilde{\phi}_{d_1\ldots d_s\,\text{v}}(x)$
and using \eqref{phid1..dsn}, we obtain
\begin{align}
  &\quad\hat{V}_{d_1\ldots d_s}(x+i\tfrac{\gamma}{2})
  +\hat{V}^*_{d_1\ldots d_s}(x-i\tfrac{\gamma}{2})
  -\tilde{\mathcal{E}}_{d_s}+\mathcal{E}_n
  \label{Vh+Vh*-Et+En=}\\
  &=\sqrt{V(x-i\tfrac{s}{2}\gamma)V^*(x-i\tfrac{s+2}{2}\gamma)}\,
  \frac{\text{W}_{\gamma}[\tilde{\phi}_{d_1},\ldots,\tilde{\phi}_{d_s}]
  (x+i\frac{\gamma}{2})}
  {\text{W}_{\gamma}[\tilde{\phi}_{d_1},\ldots,\tilde{\phi}_{d_s}]
  (x-i\frac{\gamma}{2})}
  \frac{\text{W}_{\gamma}[\tilde{\phi}_{d_1},\ldots,\tilde{\phi}_{d_s},
  \phi_n](x-i\gamma)}
  {\text{W}_{\gamma}[\tilde{\phi}_{d_1},\ldots,\tilde{\phi}_{d_s},\phi_n](x)}\n
  &\quad+\sqrt{V^*(x+i\tfrac{s}{2}\gamma)V(x+i\tfrac{s+2}{2}\gamma)}\,
  \frac{\text{W}_{\gamma}[\tilde{\phi}_{d_1},\ldots,\tilde{\phi}_{d_s}]
  (x-i\frac{\gamma}{2})}
  {\text{W}_{\gamma}[\tilde{\phi}_{d_1},\ldots,\tilde{\phi}_{d_s}]
  (x+i\frac{\gamma}{2})}
  \frac{\text{W}_{\gamma}[\tilde{\phi}_{d_1},\ldots,\tilde{\phi}_{d_s},
  \phi_n](x+i\gamma)}
  {\text{W}_{\gamma}[\tilde{\phi}_{d_1},\ldots,\tilde{\phi}_{d_s},\phi_n](x)},
  \nonumber
\end{align}
and a similar equation for the virtual state solution.

The deformed Hamiltonian $\mathcal{H}_{d_1\ldots d_s}$ can be
rewritten in the standard form:
\begin{align}
  &\mathcal{H}_{d_1\ldots d_s}
  =\mathcal{A}_{d_1\ldots d_s}^{\dagger}\mathcal{A}_{d_1\ldots d_s},
  \label{sstanham1}\\
  &\mathcal{A}_{d_1\ldots d_s}\eqdef
  i\bigl(e^{\frac{\gamma}{2}p}\sqrt{V_{d_1\ldots d_s}^*(x)}
  -e^{-\frac{\gamma}{2}p}\sqrt{V_{d_1\ldots d_s}(x)}\,\bigr),\n
  &\mathcal{A}_{d_1\ldots d_s}^{\dagger}\eqdef
  -i\bigl(\sqrt{V_{d_1\ldots d_s}(x)}\,e^{\frac{\gamma}{2}p}
  -\sqrt{V_{d_1\ldots d_s}^*(x)}\,e^{-\frac{\gamma}{2}p}\bigr),\\
  &V_{d_1\ldots d_s}(x)\eqdef
  \sqrt{V(x-i\tfrac{s}{2}\gamma)V^*(x-i\tfrac{s+2}{2}\gamma)}\n
  &\phantom{V_{d_1\ldots d_s}(x)\eqdef}\times
  \frac{\text{W}_{\gamma}[\tilde{\phi}_{d_1},\ldots,\tilde{\phi}_{d_s}]
  (x+i\frac{\gamma}{2})}
  {\text{W}_{\gamma}[\tilde{\phi}_{d_1},\ldots,\tilde{\phi}_{d_s}]
  (x-i\frac{\gamma}{2})}\,
  \frac{\text{W}_{\gamma}[\tilde{\phi}_{d_1},\ldots,\tilde{\phi}_{d_s},
  \phi_0](x-i\gamma)}
  {\text{W}_{\gamma}[\tilde{\phi}_{d_1},\ldots,\tilde{\phi}_{d_s},\phi_0](x)},
  \label{Vd1..ds}
\end{align}
in which the $\mathcal{A}$ operator annihilates the groundstate,
\begin{equation}
  \mathcal{A}_{d_1\ldots d_s}\phi_{d_1\ldots d_s\,0}(x)=0.
\end{equation}
The conditions for the equality of \eqref{sstanham1} and \eqref{Hd1..ds}
are
\begin{align}
  &V_{d_1\ldots d_s}(x)V^*_{d_1\ldots d_s}(x-i\gamma)
  =\hat{V}_{d_1\ldots d_s}(x-i\tfrac{\gamma}{2})
  \hat{V}^*_{d_1\ldots d_s}(x-i\tfrac{\gamma}{2}),\n
  &V_{d_1\ldots d_s}(x)+V^*_{d_1\ldots d_s}(x)
  =\hat{V}_{d_1\ldots d_s}(x+i\tfrac{\gamma}{2})
  +\hat{V}^*_{d_1\ldots d_s}(x-i\tfrac{\gamma}{2})
  -\tilde{\mathcal{E}}_{d_s}.
  \label{V+V*=Vh+Vh*-tE}
\end{align}
The first equation is trivially satisfied and the second equation is a
consequence of \eqref{Vh+Vh*-Et+En=} with $n=0$.

It should be stressed that the above results after $s$-deletions are
independent of the orders of deletions ($\phi_{d_1\ldots d_s\,n}(x)$
and $\tilde{\phi}_{d_1\ldots d_s\,\text{v}}(x)$ may change sign).

In order that this deformed system is well-defined {\it i.e.}
Hamiltonian $\mathcal{H}_{d_1\ldots d_s}$ is hermitian, we have to
study the singularities of $V_{d_1\ldots d_s}(x)$ and
$\phi_{d_1\ldots d_s\,n}(x)$.
We will do this for explicit examples in section \S\ref{sec:analherm}.

\section{Multi-indexed Wilson and Askey-Wilson Polynomials}
\label{sec:miop}
\setcounter{equation}{0}

In this section we apply the method of virtual states deletion to
the exactly solvable systems whose eigenstates are described by
the Wilson (W) and Askey-Wilson (AW) polynomials.
We delete $M$ virtual states labeled by
\begin{equation}
  \mathcal{D}=\{d_1,d_2,\ldots,d_M\}
  \ \ (d_j\in\mathcal{V} : \text{mutually distinct}),
\end{equation}
and denote $\mathcal{H}_{d_1\ldots d_M}$, $\phi_{d_1\ldots d_M\,n}$,
$\mathcal{A}_{d_1\ldots d_M}$, etc. simply by $\mathcal{H}_{\mathcal{D}}$,
$\phi_{\mathcal{D}\,n}$, $\mathcal{A}_{\mathcal{D}}$, etc.

We follow the notation of \cite{os13}.
Various quantities depend on a set of parameters
$\bm{\lambda}=(\lambda_1,\lambda_2,\ldots)$.

\subsection{Original Wilson and Askey-Wilson systems}
\label{sec:org_AW}

Let us consider the Wilson and Askey-Wilson cases.
Various parameters are
\begin{alignat}{2}
  \text{W}:\ \ &x_1=0,\ x_2=\infty,\ \gamma=1,
  &\ \ \bm{\lambda}=(a_1,a_2,a_3,a_4),
  &\ \ \bm{\delta}=(\tfrac12,\tfrac12,\tfrac12,\tfrac12),
  \ \ \kappa=1,\n
  \text{AW}:\ \ &x_1=0,\ x_2=\pi,\ \gamma=\log q,
  &\ \ q^{\bm{\lambda}}=(a_1,a_2,a_3,a_4),
  &\ \ \bm{\delta}=(\tfrac12,\tfrac12,\tfrac12,\tfrac12),
  \ \ \kappa=q^{-1},
\end{alignat}
where $q^{\bm{\lambda}}$ stands for
$q^{(\lambda_1,\lambda_2,\ldots)}=(q^{\lambda_1},q^{\lambda_2},\ldots)$
and $0<q<1$.
The parameters are restricted by
\begin{equation}
  \{a_1^*,a_2^*,a_3^*,a_4^*\}=\{a_1,a_2,a_3,a_4\}\ \ (\text{as a set});\quad
  \text{W}:\ \text{Re}\,a_i>0,\quad
  \text{AW}:\ |a_i|<1.
  \label{rangeorg}
\end{equation}
Here are the fundamental data \cite{os13}:
\begin{align}
  &V(x;\bm{\lambda})=\left\{
  \begin{array}{ll}
  \bigl(2ix(2ix+1)\bigr)^{-1}\prod_{j=1}^4(a_j+ix)&:\text{W}\\[2pt]
  \bigl((1-e^{2ix})(1-qe^{2ix})\bigr)^{-1}\prod_{j=1}^4(1-a_je^{ix})
  &:\text{AW}
  \end{array}\right.,
  \label{Vform}\\
  &\eta(x)=\left\{
  \begin{array}{ll}
  x^2&:\text{W}\\
  \cos x&:\text{AW}
  \end{array}\right.,\quad
  \varphi(x)=\left\{
  \begin{array}{ll}
  2x&:\text{W}\\
  2\sin x&:\text{AW}
  \end{array}\right.,
  \label{etadef}\\
  &\mathcal{E}_n(\bm{\lambda})=\left\{
  \begin{array}{ll}
  n(n+b_1-1)&:\text{W}\\[2pt]
  (q^{-n}-1)(1-b_4q^{n-1})&:\text{AW}
  \end{array}\right.,\quad
  \begin{array}{l}
  b_1\eqdef a_1+a_2+a_3+a_4,\\
  b_4\eqdef a_1a_2a_3a_4,
  \end{array}\\
  &\phi_n(x;\bm{\lambda})
  =\phi_0(x;\bm{\lambda})\check{P}_n(x;\bm{\lambda}),
  \label{factphin}\\
  &\check{P}_n(x;\bm{\lambda})=P_n\bigl(\eta(x);\bm{\lambda}\bigr)
  =\left\{\begin{array}{ll}
  W_n(\eta(x);a_1,a_2,a_3,a_4)\!\!&:\text{W}\\[2pt]
  p_n(\eta(x);a_1,a_2,a_3,a_4|q)\!\!&:\text{AW}
  \end{array}\right.\n
  &\phantom{\check{P}_n(x;\bm{\lambda})}=\left\{\begin{array}{ll}
  {\displaystyle
  (a_1+a_2)_n(a_1+a_3)_n(a_1+a_4)_n}\\[2pt]
  {\displaystyle
  \quad\times
  {}_4F_3\Bigl(
  \genfrac{}{}{0pt}{}{-n,\,n+b_1-1,\,a_1+ix,\,a_1-ix}
  {a_1+a_2,\,a_1+a_3,\,a_1+a_4}\Bigm|1\Bigr)
  }&:\text{W}\\[8pt]
  {\displaystyle
  a_1^{-n}(a_1a_2,a_1a_3,a_1a_4\,;q)_n}\\[2pt]
  {\displaystyle
  \quad\times
  {}_4\phi_3\Bigl(\genfrac{}{}{0pt}{}{q^{-n},\,b_4q^{n-1},\,
  a_1e^{ix},\,a_1e^{-ix}}{a_1a_2,\,a_1a_3,\,a_1a_4}\!\!\Bigm|\!q\,;q\Bigr)
  }&:\text{AW}
  \end{array}\right.
  \label{Pn=W,AW}\\[2pt]
  &\phantom{\check{P}_n(x;\bm{\lambda})}=
  c_n(\bm{\lambda})\eta(x)^n+(\text{lower order terms}),\n
  &c_n(\bm{\lambda})=\left\{
  \begin{array}{ll}
  (-1)^n(n+b_1-1)_n&:\text{W}\\
  2^n(b_4q^{n-1};q)_n&:\text{AW}
  \end{array}\right.,
  \label{cn}\\
  &\phi_0(x;\bm{\lambda})=\left\{
  \begin{array}{ll}
  \sqrt{(\Gamma(2ix)\Gamma(-2ix))^{-1}\prod_{j=1}^4
  \Gamma(a_j+ix)\Gamma(a_j-ix)}&:\text{W}\\[4pt]
  \sqrt{(e^{2ix}\,;q)_{\infty}(e^{-2ix}\,;q)_{\infty}
  \prod_{j=1}^4(a_je^{ix}\,;q)_{\infty}^{-1}
  (a_je^{-ix}\,;q)_{\infty}^{-1}}
  &:\text{AW}
  \end{array}\right.,
  \label{phi0=W,AW}\\
  &h_n(\bm{\lambda})=\left\{
  \begin{array}{ll}
  2\pi n!\,(n+b_1-1)_n\prod_{1\leq i<j\leq 4}\Gamma(n+a_i+a_j)\cdot
  \Gamma(2n+b_1)^{-1}&:\text{W}\\[6pt]
  2\pi(b_4q^{n-1};q)_n(b_4q^{2n};q)_{\infty}(q^{n+1};q)_{\infty}^{-1}
  \prod_{1\leq i<j\leq 4}(a_ia_jq^n;q)_{\infty}^{-1}&:\text{AW}
  \end{array}\right..
  \label{hn}
\end{align}
Here $W_n$ and $p_n$ in \eqref{Pn=W,AW} are the Wilson and the Askey-Wilson
polynomials \cite{koeswart} and the symbols $(a)_n$ and $(a;q)_n$ are
($q$-)shifted factorials.
We have $\phi^*_0(x;\bm{\lambda})=\phi_0(x;\bm{\lambda})$
and $\check{P}^*_n(x;\bm{\lambda})=\check{P}_n(x;\bm{\lambda})$.
Note that
\begin{align}
  &\phi_0(x;\bm{\lambda}+\bm{\delta})
  =\varphi(x)\sqrt{V(x+i\tfrac{\gamma}{2};\bm{\lambda})}\,
  \phi_0(x+i\tfrac{\gamma}{2};\bm{\lambda}),
  \label{phi0(l+d)}\\
  &V(x;\bm{\lambda}+\bm{\delta})
  =\kappa^{-1}\frac{\varphi(x-i\gamma)}{\varphi(x)}
  V(x-i\tfrac{\gamma}{2};\bm{\lambda}).
  \label{varphiprop3}
\end{align}
The sinusoidal coordinate $\eta(x)$ has a special dynamical meaning
\cite{os12,os13,os7}. The Heisenberg operator solution for $\eta(x)$
can be expressed in a closed form and its time evolution is a
sinusoidal motion.
The hermiticity of the Hamiltonian is satisfied because the function
$V\phi_0^2$ has the property
\begin{equation}
  \eqref{rangeorg}\Leftrightarrow
  \text{$V(x+i\tfrac{\gamma}{2};\bm{\lambda})
  \phi_0(x+i\tfrac{\gamma}{2};\bm{\lambda})^2$ has no poles in the
  rectangular domain $D_{\gamma}$},
  \label{Vphi0^2}
\end{equation}
and the function $G$ \eqref{Gdef}
($\check{\mathcal{R}}(x)=\check{P}_n(x;\bm{\lambda})$) satisfies
$G(x_1+ix)=G^*(x_1-ix)$, $G(x_2+ix)=0=G^*(x_2-ix)$ for W and
$G(x_1+ix)=G^*(x_1-ix)$, $G(x_2+ix)=G^*(x_2-ix)$ for AW.
Note that eq.\eqref{phi0(l+d)} implies
\begin{equation}
  V(x+i\tfrac{\gamma}{2};\bm{\lambda})
  \phi_0(x+i\tfrac{\gamma}{2};\bm{\lambda})^2
  =\frac{\phi_0(x;\bm{\lambda}+\bm{\delta})^2}{\varphi(x)^2}.
\end{equation}

The system is shape invariant \cite{genden,os13},
\begin{equation}
  \mathcal{A}(\bm{\lambda})\mathcal{A}(\bm{\lambda})^{\dagger}
  =\kappa\mathcal{A}(\bm{\lambda}+\bm{\delta})^{\dagger}
  \mathcal{A}(\bm{\lambda}+\bm{\delta})+\mathcal{E}_1(\bm{\lambda}),
\end{equation}
which is a sufficient condition for exact solvability and it provides
the explicit formulas for the energy eigenvalues and the eigenfunctions,
{\em i.e.} the generalised Rodrigues formulas \cite{os13}.
The action of the operators $\mathcal{A}(\bm{\lambda})$ and
$\mathcal{A}(\bm{\lambda})^{\dagger}$ on the eigenfunctions is
\begin{equation}
  \mathcal{A}(\bm{\lambda})\phi_n(x;\bm{\lambda})
  =f_n(\bm{\lambda})
  \phi_{n-1}\bigl(x;\bm{\lambda}+\bm{\delta}\bigr),\quad
  \mathcal{A}(\bm{\lambda})^{\dagger}
  \phi_{n-1}\bigl(x;\bm{\lambda}+\bm{\delta}\bigr)
  =b_{n-1}(\bm{\lambda})\phi_n(x;\bm{\lambda}).
  \label{Aphi=,Adphi=}
\end{equation}
The factors of the energy eigenvalue, $f_n(\bm{\lambda})$ and
$b_{n-1}(\bm{\lambda})$,
$\mathcal{E}_n(\bm{\lambda})=f_n(\bm{\lambda})b_{n-1}(\bm{\lambda})$,
are given by
\begin{equation}
  f_n(\bm{\lambda})\eqdef\left\{
  \begin{array}{ll}
  -n(n+b_1-1)&:\text{W}\\
  q^{\frac{n}{2}}(q^{-n}-1)(1-b_4q^{n-1})&:\text{AW}
  \end{array}\right.,
  \quad
  b_{n-1}(\bm{\lambda})\eqdef\left\{
  \begin{array}{ll}
  -1&:\text{W}\\
  q^{-\frac{n}{2}}&:\text{AW}
  \end{array}\right..
\end{equation}
The forward and backward shift operators $\mathcal{F}(\bm{\lambda})$ and
$\mathcal{B}(\bm{\lambda})$ are defined by
\begin{align}
  &\mathcal{F}(\bm{\lambda})\eqdef
  \phi_0(x;\bm{\lambda}+\bm{\delta})^{-1}\circ
  \mathcal{A}(\bm{\lambda})\circ\phi_0(x;\bm{\lambda})
  =i\varphi(x)^{-1}(e^{\frac{\gamma}{2}p}-e^{-\frac{\gamma}{2}p}),
  \label{Fdef}\\
  &\mathcal{B}(\bm{\lambda})\eqdef
  \phi_0(x;\bm{\lambda})^{-1}\circ
  \mathcal{A}(\bm{\lambda})^{\dagger}
  \circ\phi_0(x;\bm{\lambda}+\bm{\delta})
  =-i\bigl(V(x;\bm{\lambda})e^{\frac{\gamma}{2}p}
  -V^*(x;\bm{\lambda})e^{-\frac{\gamma}{2}p}\bigr)\varphi(x),
  \label{Bdef}
\end{align}
and their action on the polynomials is
\begin{equation}
  \mathcal{F}(\bm{\lambda})\check{P}_n(x;\bm{\lambda})
  =f_n(\bm{\lambda})\check{P}_{n-1}(x;\bm{\lambda}+\bm{\delta}),\quad
  \mathcal{B}(\bm{\lambda})\check{P}_{n-1}(x;\bm{\lambda}+\bm{\delta})
  =b_{n-1}(\bm{\lambda})\check{P}_n(x;\bm{\lambda}).
  \label{FP=,BP=}
\end{equation}
The second order difference operator $\widetilde{\mathcal{H}}(\bm{\lambda})$
acting on the polynomial eigenfunctions is square root free. It is defined by
\begin{align}
  &\widetilde{\mathcal{H}}(\bm{\lambda})\eqdef
  \phi_0(x;\bm{\lambda})^{-1}\circ\mathcal{H}(\bm{\lambda})
  \circ\phi_0(x;\bm{\lambda})
  =\mathcal{B}(\bm{\lambda})\mathcal{F}(\bm{\lambda})\n
  &\phantom{\widetilde{\mathcal{H}}_{\ell}(\bm{\lambda})}
  =V(x;\bm{\lambda})(e^{\gamma p}-1)
  +V^*(x;\bm{\lambda})(e^{-\gamma p}-1),\\
  &\widetilde{\mathcal{H}}(\bm{\lambda})\check{P}_n(x;\bm{\lambda})
  =\mathcal{E}_n(\bm{\lambda})\check{P}_n(x;\bm{\lambda}).
  \label{HtP=EP}
\end{align}

\subsection{Discrete symmetries of the Wilson and Askey-Wilson systems}
\label{sec:discr}

Since the potential function $V(x;\bm{\lambda})$ \eqref{Vform} is
invariant under the permutation of $(a_1,a_2,a_3,a_4)$, the system is
symmetric in $(a_1,a_2,a_3,a_4)$.
So are the groundstate wavefunction $\phi_0(x;\bm{\lambda})$
\eqref{phi0=W,AW} and the eigenpolynomial $\check{P}_n(x;\bm{\lambda})$
\eqref{Pn=W,AW}.

In the following we restrict the parameters as follows:
\begin{equation}
  a_1,a_2\in\mathbb{R}\ \ \text{or}\ \ a_2^*=a_1\ ;\quad
  a_3,a_4\in\mathbb{R}\ \ \text{or}\ \ a_4^*=a_3.
\end{equation}
Let us introduce twist operations $\mathfrak{t}^{\I}$, $\mathfrak{t}^{\II}$
and constants $\tilde{\bm{\delta}}^{\I}$, $\tilde{\bm{\delta}}^{\II}$,
\begin{align}
  \mathfrak{t}^{\I}(\bm{\lambda})
  &\eqdef(1-\lambda_1,1-\lambda_2,\lambda_3,\lambda_4),\,\quad
  \tilde{\bm{\delta}}^{\I}\eqdef(-\tfrac12,-\tfrac12,\tfrac12,\tfrac12),\n
  \mathfrak{t}^{\II}(\bm{\lambda})
  &\eqdef(\lambda_1,\lambda_2,1-\lambda_3,1-\lambda_4),\quad
  \tilde{\bm{\delta}}^{\II}\eqdef(\tfrac12,\tfrac12,-\tfrac12,-\tfrac12).
  \label{twist}
\end{align}
By using these twist operations, we define deformed potential functions
$V'$ by
\begin{equation}
  V^{\prime\,\I}(x;\bm{\lambda})\eqdef
  V\bigl(x;\mathfrak{t}^{\I}(\bm{\lambda})\bigr),\quad
  V^{\prime\,\II}(x;\bm{\lambda})\eqdef
  V\bigl(x;\mathfrak{t}^{\II}(\bm{\lambda})\bigr).
  \label{V'}
\end{equation}
These $V'$s satisfy \eqref{propV'} with
\begin{align}
  &\alpha^{\I}(\bm{\lambda})=\left\{
  \begin{array}{ll}
  1&:\text{W}\\
  a_1a_2q^{-1}&:\text{AW}
  \end{array}\right.,\quad
  \alpha^{\prime\,\I}(\bm{\lambda})=\left\{
  \begin{array}{ll}
  -(a_1+a_2-1)(a_3+a_4)&:\text{W}\\
  -(1-a_1a_2q^{-1})(1-a_3a_4)&:\text{AW}
  \end{array}\right.,\n
  &\alpha^{\II}(\bm{\lambda})=\left\{
  \begin{array}{ll}
  1&:\text{W}\\
  a_3a_4q^{-1}&:\text{AW}
  \end{array}\right.,\quad
  \alpha^{\prime\,\II}(\bm{\lambda})=\left\{
  \begin{array}{ll}
  -(a_3+a_4-1)(a_1+a_2)&:\text{W}\\
  -(1-a_3a_4q^{-1})(1-a_1a_2)&:\text{AW}
  \end{array}\right..
\end{align}
(For $\alpha>0$, we have $a_1a_2>0$ and $a_3a_4>0$ for AW case.
For $\alpha'<0$, we will restrict the parameters further as in
\eqref{range2}.)
We obtain a linear relation between the two Hamiltonians
\eqref{H=aH'+a'}. The virtual state wavefunctions satisfying
\eqref{Hphiv=} are given by
\begin{align}
  &\tilde{\phi}^{\I}_0(x;\bm{\lambda})\eqdef
  \phi_0\bigl(x;\mathfrak{t}^{\I}(\bm{\lambda})\bigr),\quad
  \tilde{\phi}^{\I}_{\text{v}}(x;\bm{\lambda})\eqdef
  \phi_{\text{v}}\bigl(x;\mathfrak{t}^{\I}(\bm{\lambda})\bigr)
  =\tilde{\phi}_0^{\I}(x;\bm{\lambda})
  \check{\xi}^{\I}_{\text{v}}(x;\bm{\lambda})
  \quad(\text{v}\in\mathcal{V}^{\I}),\n
  &\check{\xi}^{\I}_{\text{v}}(x;\bm{\lambda})\eqdef
  \xi^{\I}_{\text{v}}\bigl(\eta(x);\bm{\lambda}\bigr)\eqdef
  \check{P}_{\text{v}}\bigl(x;\mathfrak{t}^{\I}(\bm{\lambda})\bigr)
  =P_{\text{v}}\bigl(\eta(x);\mathfrak{t}^{\I}(\bm{\lambda})\bigr),\n
  &\tilde{\phi}^{\II}_0(x;\bm{\lambda})\eqdef
  \phi_0\bigl(x;\mathfrak{t}^{\II}(\bm{\lambda})\bigr),\quad
  \tilde{\phi}^{\II}_{\text{v}}(x;\bm{\lambda})\eqdef
  \phi_{\text{v}}\bigl(x;\mathfrak{t}^{\II}(\bm{\lambda})\bigr)
  =\tilde{\phi}_0^{\II}(x;\bm{\lambda})
  \check{\xi}^{\II}_{\text{v}}(x;\bm{\lambda})
  \quad(\text{v}\in\mathcal{V}^{\II}),\n
  &\check{\xi}^{\II}_{\text{v}}(x;\bm{\lambda})\eqdef
  \xi^{\II}_{\text{v}}\bigl(\eta(x);\bm{\lambda}\bigr)\eqdef
  \check{P}_{\text{v}}\bigl(x;\mathfrak{t}^{\II}(\bm{\lambda})\bigr)
  =P_{\text{v}}\bigl(\eta(x);\mathfrak{t}^{\II}(\bm{\lambda})\bigr).
  \label{xiv=}
\end{align}
The virtual state polynomials $\xi_{\text{v}}(\eta;\bm{\lambda})$
are polynomials of degree $\text{v}$ in $\eta$.
They are chosen `real,'
$\tilde{\phi}^*_0(x;\bm{\lambda})=\tilde{\phi}_0(x;\bm{\lambda})$,
$\check{\xi}^*_{\text{v}}(x;\bm{\lambda})
=\check{\xi}_{\text{v}}(x;\bm{\lambda})$
and the virtual energies are $\mathcal{E}'_{\text{v}}(\bm{\lambda})
=\mathcal{E}_{\text{v}}\bigl(\mathfrak{t}(\bm{\lambda})\bigr)$:
\begin{align}
  &\tilde{\mathcal{E}}^{\I}_{\text{v}}(\bm{\lambda})=\left\{
  \begin{array}{ll}
  -(a_1+a_2-\text{v}-1)(a_3+a_4+\text{v})&:\text{W}\\
  -(1-a_1a_2q^{-\text{v}-1})(1-a_3a_4q^{\text{v}})&:\text{AW}
  \end{array}\right.,\n
  &\tilde{\mathcal{E}}^{\II}_{\text{v}}(\bm{\lambda})=\left\{
  \begin{array}{ll}
  -(a_3+a_4-\text{v}-1)(a_1+a_2+\text{v})&:\text{W}\\
  -(1-a_3a_4q^{-\text{v}-1})(1-a_1a_2q^{\text{v}})&:\text{AW}
  \end{array}\right..
  \label{tEv}
\end{align}
Note that $\alpha'(\bm{\lambda})=\tilde{\mathcal{E}}_0(\bm{\lambda})<0$ and
\begin{alignat}{2}
  \text{W}:\ \ &\tilde{\mathcal{E}}^{\I}_{\text{v}}(\bm{\lambda})<0
  \ \Leftrightarrow\ a_1+a_2>\text{v}+1,&\quad
  &\tilde{\mathcal{E}}^{\II}_{\text{v}}(\bm{\lambda})<0
  \ \Leftrightarrow\ a_3+a_4>\text{v}+1,\n
  \text{AW}:\ \ &\tilde{\mathcal{E}}^{\I}_{\text{v}}(\bm{\lambda})<0
  \ \Leftrightarrow\ 0<a_1a_2<q^{\text{v}+1},&\quad
  &\tilde{\mathcal{E}}^{\II}_{\text{v}}(\bm{\lambda})<0
  \ \Leftrightarrow\ 0<a_3a_4<q^{\text{v}+1},
  \label{tEv<0}
\end{alignat}
for $\text{v}\geq 0$.
We choose $\mathcal{V}^{\I}$ and $\mathcal{V}^{\II}$ as
\begin{equation}
  \mathcal{V}^{\I}=\bigl\{1,2,\ldots,[\lambda_1+\lambda_2-1]'\bigr\},\quad
  \mathcal{V}^{\II}=\bigl\{1,2,\ldots,[\lambda_3+\lambda_4-1]'\bigr\},
  \label{vrange}
\end{equation}
where $[x]'$ denotes the greatest integer not equal or exceeding $x$.
We will not use the label 0 states for deletion,
see \eqref{dM1=0}--\eqref{d'M2=0}.

For later use, we define the following functions
(recall $x^{(n)}_j$ in \eqref{Wdef}):
\begin{align}
  &\nu^{\I}(x;\bm{\lambda})\eqdef
  \frac{\phi_0(x;\bm{\lambda})}{\tilde{\phi}^{\I}_0(x;\bm{\lambda})},\quad
  r_j^{\I}(x^{(M)}_j;\bm{\lambda},M)\eqdef
  \frac{\nu^{\I}(x^{(M)}_j;\bm{\lambda})}
  {\nu^{\I}\bigl(x;\bm{\lambda}+(M-1)\tilde{\bm{\delta}}^{\I}\bigr)}
  \ \ (1\leq j\leq M),\n
  &\nu^{\II}(x;\bm{\lambda})\eqdef
  \frac{\phi_0(x;\bm{\lambda})}{\tilde{\phi}^{\II}_0(x;\bm{\lambda})},\quad
  r_j^{\II}(x^{(M)}_j;\bm{\lambda},M)\eqdef
  \frac{\nu^{\II}(x^{(M)}_j;\bm{\lambda})}
  {\nu^{\II}\bigl(x;\bm{\lambda}+(M-1)\tilde{\bm{\delta}}^{\II}\bigr)}
  \ \ (1\leq j\leq M).
\end{align}
Their explicit forms are
\begin{align}
  r^{\I}_j(x^{(M)}_j;\bm{\lambda},M)
  &=\alpha^{\I}\bigl(\bm{\lambda}
  +(M-1)\tilde{\bm{\delta}}^{\I}\bigr)^{-\frac12(M-1)}
  \kappa^{\frac12(M-1)^2-(j-1)(M-j)}\\
  &\quad\times\left\{
  \begin{array}{ll}
  {\displaystyle
  \prod_{k=1,2}(a_k-\tfrac{M-1}{2}+ix)_{j-1}(a_k-\tfrac{M-1}{2}-ix)_{M-j}
  }&:\text{W}\\
  {\displaystyle
  e^{ix(M+1-2j)}\prod_{k=1,2}(a_kq^{-\frac{M-1}{2}}e^{ix};q)_{j-1}
  (a_kq^{-\frac{M-1}{2}}e^{-ix};q)_{M-j}
  }&:\text{AW}
  \end{array}\right.,\n
  r^{\II}_j(x^{(M)}_j;\bm{\lambda},M)
  &=\alpha^{\II}\bigl(\bm{\lambda}
  +(M-1)\tilde{\bm{\delta}}^{\II}\bigr)^{-\frac12(M-1)}
  \kappa^{\frac12(M-1)^2-(j-1)(M-j)}\\
  &\quad\times\left\{
  \begin{array}{ll}
  {\displaystyle
  \prod_{k=3,4}(a_k-\tfrac{M-1}{2}+ix)_{j-1}(a_k-\tfrac{M-1}{2}-ix)_{M-j}
  }&:\text{W}\\
  {\displaystyle
  e^{ix(M+1-2j)}\prod_{k=3,4}(a_kq^{-\frac{M-1}{2}}e^{ix};q)_{j-1}
  (a_kq^{-\frac{M-1}{2}}e^{-ix};q)_{M-j}
  }&:\text{AW}
  \end{array}\right..\nonumber
\end{align}
The auxiliary function $\varphi_M(x)$ \cite{gos} is defined by:
\begin{align}
  \varphi_M(x)&\eqdef
  \varphi(x)^{[\frac{M}{2}]}\prod_{k=1}^{M-2}
  \bigl(\varphi(x-i\tfrac{k}{2}\gamma)\varphi(x+i\tfrac{k}{2}\gamma)
  \bigr)^{[\frac{M-k}{2}]}\n
  &=\prod_{1\leq j<k\leq M}
  \frac{\eta(x^{(M)}_j)-\eta(x^{(M)}_k)}
  {\varphi(i\frac{j}{2}\gamma)}
  \times\left\{
  \begin{array}{ll}
  1&:\text{W}\\
  (-2)^{\frac12M(M-1)}&:\text{AW}
  \end{array}\right.,
  \label{varphiMdef}
\end{align}
and $\varphi_0(x)=\varphi_1(x)=1$.
Here $[x]$ denotes the greatest integer not exceeding $x$.

\subsection{Explicit Forms of Multi-indexed Wilson and Askey-Wilson
polynomials}
\label{sec:miop_AW}

We delete $M=M_{\I}+M_{\II}$ virtual states
$\mathcal{D}=\{d^{\I}_1,\ldots,d^{\I}_{M_{\I}},
d^{\II}_1,\ldots,d^{\II}_{M_{\II}}\}$, where $d^{\I}_j$ and $d^{\II}_j$
are labels of the type $\I$ and $\II$ virtual states, respectively.
We restrict the parameter range further as follows:
\begin{align}
  \text{W}:\ \ &\text{Re}\,a_i>\tfrac12(\max_j\{d^{\I}_j\}+1)\ \ (i=1,2),
  \ \ \text{Re}\,a_i>\tfrac12(\max_j\{d^{\II}_j\}+1)\ \ (i=3,4),
  \label{range2}\\
  \text{AW}:\ \ &|a_i|<q^{\frac12(\max_j\{d^{\I}_j\}+1)}\ (i=1,2),
  \ \ |a_i|<q^{\frac12(\max_j\{d^{\II}_j\}+1)}\ (i=3,4),
  \ \ a_1a_2,a_3a_4>0.
  \nonumber
\end{align}
Then the condition $\tilde{\mathcal{E}}_{\text{v}}(\bm{\lambda})<0$
($\text{v}\in\mathcal{D}$) is satisfied, see \eqref{tEv<0}.
We assume that the parameters are so chosen that
$\check{\xi}_{\text{v}}(x;\bm{\lambda})\neq 0$ for $x_1\leq x\leq x_2$.
For example, if we take the parameters as
\begin{align*}
  \circ\ \ &a_i\in\mathbb{R},
  \ \ -1<\lambda_j-\lambda_k<1\ \ (j=1,2; k=3,4),\\
  \circ\ \ &a_1,a_2\in\mathbb{R},\ a_4^*=a_3,
  \ \ \lambda_1+\lambda_2+2\text{v}<\lambda_3+\lambda_4,\\
  \circ\ \ &a_3,a_4\in\mathbb{R},\ a_2^*=a_1,
  \ \ \lambda_3+\lambda_4+2\text{v}<\lambda_1+\lambda_2,
\end{align*}
then $\check{\xi}_{\text{v}}(x;\bm{\lambda})$ and
$\check{\xi}_{\text{v}}(x;\bm{\lambda}+\bm{\delta})$ have a definite sign
for real $x$.

Let us write down $\phi_{\mathcal{D}\,n}$ \eqref{phid1..dsn} concretely.
The Casoratians in \eqref{phid1..dsn} are reduced to the following determinants,
by which we define two polynomials $\check{\Xi}_{\mathcal{D}}(x;\bm{\lambda})$
and $\check{P}_{\mathcal{D},n}(x;\bm{\lambda})$, to be called the denominator
polynomial and the multi-indexed orthogonal polynomial, respectively:
\begin{align}
  &i^{\frac12M(M-1)}\left|
  \begin{array}{llllll}
  \vec{X}^{(M)}_{d^{\I}_1}&\cdots&\vec{X}^{(M)}_{d^{\I}_{M_{\I}}}&
  \vec{Y}^{(M)}_{d^{\II}_1}&\cdots&\vec{Y}^{(M)}_{d^{\II}_{M_{\II}}}\\
  \end{array}\right|
  =\varphi_M(x)\check{\Xi}_{\mathcal{D}}(x;\bm{\lambda})\times A,\n
  &\quad A=\left\{
  \begin{array}{ll}
  \prod_{k=3,4}\prod_{j=1}^{M_{\I}-1}
  (a_k-\frac{M-1}{2}+ix,a_k-\frac{M-1}{2}-ix)_j\\[4pt]
  \ \ \times\prod_{k=1,2}\prod_{j=1}^{M_{\II}-1}
  (a_k-\frac{M-1}{2}+ix,a_k-\frac{M-1}{2}-ix)_j
  &:\text{W}\\[4pt]
  \prod_{k=3,4}\prod_{j=1}^{M_{\I}-1}a_k^{-j}q^{\frac14j(j+1)}
  (a_kq^{-\frac{M-1}{2}}e^{ix},a_kq^{-\frac{M-1}{2}}e^{-ix};q)_j\\[4pt]
  \ \ \times\prod_{k=1,2}\prod_{j=1}^{M_{\II}-1}a_k^{-j}q^{\frac14j(j+1)}
  (a_kq^{-\frac{M-1}{2}}e^{ix},a_kq^{-\frac{M-1}{2}}e^{-ix};q)_j
  &:\text{AW}
  \end{array}\right.,
  \label{cXiDdef}\\[2pt]
  &i^{\frac12M(M+1)}\left|
  \begin{array}{lllllll}
  \vec{X}^{(M+1)}_{d^{\I}_1}&\cdots&\vec{X}^{(M+1)}_{d^{\I}_{M_{\I}}}&
  \vec{Y}^{(M+1)}_{d^{\II}_1}&\cdots&\vec{Y}^{(M+1)}_{d^{\II}_{M_{\II}}}&
  \vec{Z}^{(M+1)}_n\\
  \end{array}\right|\n
  &=\varphi_{M+1}(x)\check{P}_{\mathcal{D},n}(x;\bm{\lambda})\times B,\n
  &\quad B=\left\{
  \begin{array}{ll}
  \prod_{k=3,4}\prod_{j=1}^{M_{\I}}
  (a_k-\frac{M}{2}+ix,a_k-\frac{M}{2}-ix)_j\\[4pt]
  \ \ \times\prod_{k=1,2}\prod_{j=1}^{M_{\II}}
  (a_k-\frac{M}{2}+ix,a_k-\frac{M}{2}-ix)_j
  &:\text{W}\\[4pt]
  \prod_{k=3,4}\prod_{j=1}^{M_{\I}}a_k^{-j}q^{\frac14j(j+1)}
  (a_kq^{-\frac{M}{2}}e^{ix},a_kq^{-\frac{M}{2}}e^{-ix};q)_j\\[4pt]
  \ \ \times\prod_{k=1,2}\prod_{j=1}^{M_{\II}}a_k^{-j}q^{\frac14j(j+1)}
  (a_kq^{-\frac{M}{2}}e^{ix},a_kq^{-\frac{M}{2}}e^{-ix};q)_j
  &:\text{AW}
  \end{array}\right.,
  \label{cPDndef}
\end{align}
where
\begin{align}
  &\bigl(\vec{X}^{(M)}_{\text{v}}\bigr)_j
  =r^{\II}_j(x^{(M)}_j;\bm{\lambda},M)
  \check{\xi}^{\I}_{\text{v}}(x^{(M)}_j;\bm{\lambda}),\qquad
  (1\leq j\leq M),\n
  &\bigl(\vec{Y}^{(M)}_{\text{v}}\bigr)_j
  =r^{\I}_j(x^{(M)}_j;\bm{\lambda},M)
  \check{\xi}^{\II}_{\text{v}}(x^{(M)}_j;\bm{\lambda}),\n
  &\bigl(\vec{Z}^{(M)}_n\bigr)_j
  =r^{\II}_j(x^{(M)}_j;\bm{\lambda},M)r^{\I}_j(x^{(M)}_j;\bm{\lambda},M)
  \check{P}_n(x^{(M)}_j;\bm{\lambda}).
\end{align}
They are `real',
$\check{\Xi}^*_{\mathcal{D}}(x;\bm{\lambda})
=\check{\Xi}_{\mathcal{D}}(x;\bm{\lambda})$ and
$\check{P}^*_{\mathcal{D},n}(x;\bm{\lambda})
=\check{P}_{\mathcal{D},n}(x;\bm{\lambda})$.
After some calculation, the eigenfunction \eqref{phid1..dsn} is rewritten as
\begin{align}
  \phi_{\mathcal{D}\,n}(x;\bm{\lambda})&=
  \alpha^{\I}(\bm{\lambda}^{[M_{\I},M_{\II}]})^{\frac12M_{\I}}
  \alpha^{\II}(\bm{\lambda}^{[M_{\I},M_{\II}]})^{\frac12M_{\II}}
  \kappa^{-\frac14M_{\I}(M_{\I}+1)-\frac14M_{\II}(M_{\II}+1)
  +\frac52M_{\I}M_{\II}}\n
  &\quad\times
  \psi_{\mathcal{D}}(x;\bm{\lambda})
  \check{P}_{\mathcal{D},n}(x;\bm{\lambda}),
  \label{phiDn}\\
  \psi_{\mathcal{D}}(x;\bm{\lambda})&\eqdef
  \frac{\phi_0(x;\bm{\lambda}^{[M_{\I},M_{\II}]})}
  {\sqrt{\check{\Xi}_{\mathcal{D}}(x-i\frac{\gamma}{2};\bm{\lambda})
  \check{\Xi}_{\mathcal{D}}(x+i\frac{\gamma}{2};\bm{\lambda})}},\quad
  \bm{\lambda}^{[M_{\I},M_{\II}]}\eqdef
  \bm{\lambda}+M_{\I}\tilde{\bm{\delta}}^{\I}+M_{\II}\tilde{\bm{\delta}}^{\II}.
  \label{psiD}
\end{align}
The ground state $\phi_{\mathcal{D}\,0}$ is annihilated by
$\mathcal{A}_{\mathcal{D}}$,
$\mathcal{A}_{\mathcal{D}}(\bm{\lambda})
\phi_{\mathcal{D}\,0}(x;\bm{\lambda})=0$.
By construction $\psi_{\mathcal{D}}(x;\bm{\lambda})$ is positive definite
in $x_1\le x\le x_2$.
By using the properties of $\eta(x)$, $r_j(x;\bm{\lambda},M)$,
$\varphi_M(x)$ and the determinants, we can show that these
$\check{\Xi}_{\mathcal{D}}(x)$ \eqref{cXiDdef} and
$\check{P}_{\mathcal{D},n}(x)$ \eqref{cPDndef} are polynomials in the
sinusoidal coordinate $\eta(x)$:
\begin{equation}
  \check{\Xi}_{\mathcal{D}}(x;\bm{\lambda})\eqdef
  \Xi_{\mathcal{D}}\bigl(\eta(x);\bm{\lambda}\bigr),
  \quad
  \check{P}_{\mathcal{D},n}(x;\bm{\lambda})\eqdef
  P_{\mathcal{D},n}\bigl(\eta(x);\bm{\lambda}\bigr),
  \label{XiP_poly}
\end{equation}
and their degrees are generically $\ell$ and $\ell+n$, respectively
(See \eqref{cXiD}--\eqref{cPDn}).
Here $\ell$ is
\begin{equation}
  \ell\eqdef\sum_{j=1}^{M_{\I}}d^{\I}_j+\sum_{j=1}^{M_{\II}}d^{\II}_j
  -\tfrac12M_{\I}(M_{\I}-1)-\tfrac12M_{\II}(M_{\II}-1)+M_{\I}M_{\II}.
  \label{lform}
\end{equation}
We assume that the parameters are so chosen that
$\check{\Xi}_{\mathcal{D}}(x;\bm{\lambda})\neq 0$ for $x_1\leq x\leq x_2$.
We have also $\check{\Xi}_{\mathcal{D}}(x;\bm{\lambda}+\bm{\delta})\neq 0$
($x_1\leq x\leq x_2$).
By using these and \eqref{xiDl->l+d}--\eqref{xiDl+d->lII}, we can show that
$\check{\Xi}_{\mathcal{D}}(x\mp i\frac{\gamma}{2};\bm{\lambda})\neq 0$
($x_1\leq x\leq x_2$).
The lowest degree multi-indexed orthogonal polynomial
$\check{P}_{\mathcal{D},0}(x;\bm{\lambda})$ is related to
$\check{\Xi}_{\mathcal{D}}(x;\bm{\lambda})$ by the parameter shift
$\bm{\lambda}\to\bm{\lambda}+\bm{\delta}$:
\begin{equation}
  \check{P}_{\mathcal{D},0}(x;\bm{\lambda})
  =A\,\check{\Xi}_{\mathcal{D}}(x;\bm{\lambda}+\bm{\delta}),
  \label{PD0=XiD}
\end{equation}
where the proportionality constant $A$ is given in \eqref{A_PD0=XiD}.
This can be shown by using \eqref{xil->l+d}--\eqref{xil+d->lII} etc.
The potential function $V_{\mathcal{D}}$ \eqref{Vd1..ds} after $M$ deletions
($s=M$) can be expressed neatly in terms of the denominator polynomial:
\begin{equation}
  V_{\mathcal{D}}(x;\bm{\lambda})
  =V(x;\bm{\lambda}^{[M_{\I},M_{\II}]})\,
  \frac{\check{\Xi}_{\mathcal{D}}(x+i\frac{\gamma}{2};\bm{\lambda})}
  {\check{\Xi}_{\mathcal{D}}(x-i\frac{\gamma}{2};\bm{\lambda})}
  \frac{\check{\Xi}_{\mathcal{D}}(x-i\gamma;\bm{\lambda}+\bm{\delta})}
  {\check{\Xi}_{\mathcal{D}}(x;\bm{\lambda}+\bm{\delta})}.
  \label{VD2}
\end{equation}
The orthogonality relation \eqref{(phid1..dsm,phid1..dsn)} is
\begin{align}
  &\int_{x_1}^{x_2}\!\!dx\,\psi_{\mathcal{D}}(x;\bm{\lambda})^2
  \check{P}_{\mathcal{D},n}(x;\bm{\lambda})
  \check{P}_{\mathcal{D},n}(x;\bm{\lambda})
  =h_{\mathcal{D},n}(\bm{\lambda})\delta_{nm}
  \ \ (n,m=0,1,2,\ldots),\n
  &\quad h_{\mathcal{D},n}(\bm{\lambda})=h_n(\bm{\lambda})
  \kappa^{\frac12M_{\I}(M_{\I}+1)+\frac12M_{\II}(M_{\II}+1)-5M_{\I}M_{\II}}
  \alpha^{\I}(\bm{\lambda}^{[M_{\I},M_{\II}]})^{-M_{\I}}
  \alpha^{\II}(\bm{\lambda}^{[M_{\I},M_{\II}]})^{-M_{\II}}\n
  &\phantom{\quad h_{\mathcal{D},n}(\bm{\lambda})=}
  \times\prod_{j=1}^{M_{\I}}\bigl(\mathcal{E}_n(\bm{\lambda})
  -\tilde{\mathcal{E}}_{d^{\I}_j}(\bm{\lambda})\bigr)\cdot
  \prod_{j=1}^{M_{\II}}\bigl(\mathcal{E}_n(\bm{\lambda})
  -\tilde{\mathcal{E}}_{d^{\II}_j}(\bm{\lambda})\bigr).
  \label{ortho}
\end{align}

The shape invariance of the original system is inherited by the
deformed systems.
The operators $\hat{\mathcal{A}}_{d_1\ldots d_{s+1}}(\bm{\lambda})$ and
$\hat{\mathcal{A}}_{d_1\ldots d_{s+1}}(\bm{\lambda})^{\dagger}$ intertwine
the two Hamiltonians $\mathcal{H}_{d_1\ldots d_s}(\bm{\lambda})$ and
$\mathcal{H}_{d_1\ldots d_{s+1}}(\bm{\lambda})$,
\begin{align}
  \hat{\mathcal{A}}_{d_1\ldots d_{s+1}}(\bm{\lambda})^{\dagger}
  \hat{\mathcal{A}}_{d_1\ldots d_{s+1}}(\bm{\lambda})
  &=\mathcal{H}_{d_1\ldots d_s}(\bm{\lambda})
  -\tilde{\mathcal{E}}_{d_{s+1}}(\bm{\lambda}),\n
  \hat{\mathcal{A}}_{d_1\ldots d_{s+1}}(\bm{\lambda})
  \hat{\mathcal{A}}_{d_1\ldots d_{s+1}}(\bm{\lambda})^{\dagger}
  &=\mathcal{H}_{d_1\ldots d_{s+1}}(\bm{\lambda})
  -\tilde{\mathcal{E}}_{d_{s+1}}(\bm{\lambda}).
\end{align}
It is important that they have no zero mode, so that the eigenstates of
the two Hamiltonians are mapped one to one.
In other words, the two Hamiltonians
$\mathcal{H}_{d_1\ldots d_s}(\bm{\lambda})$ and
$\mathcal{H}_{d_1\ldots d_{s+1}}(\bm{\lambda})$ are exactly isospectral.
By the same argument given in \S\,4 of \cite{os20}, the shape invariance
of $\mathcal{H}(\bm{\lambda})$ is inherited by
$\mathcal{H}_{d_1}(\bm{\lambda})$, $\mathcal{H}_{d_1d_2}(\bm{\lambda})$,
$\cdots$.
Therefore the Hamiltonian $\mathcal{H}_{\mathcal{D}}(\bm{\lambda})$
is shape invariant:
\begin{equation}
  \mathcal{A}_{\mathcal{D}}(\bm{\lambda})
  \mathcal{A}_{\mathcal{D}}(\bm{\lambda})^{\dagger}
  =\kappa\mathcal{A}_{\mathcal{D}}(\bm{\lambda}+\bm{\delta})^{\dagger}
  \mathcal{A}_{\mathcal{D}}(\bm{\lambda}+\bm{\delta})
  +\mathcal{E}_1(\bm{\lambda}).
  \label{shapeinvD}
\end{equation}
As a consequence of the shape invariance, the actions of
$\mathcal{A}_{\mathcal{D}}(\bm{\lambda})$ and
$\mathcal{A}_{\mathcal{D}}(\bm{\lambda})^{\dagger}$ on the eigenfunctions
$\phi_{\mathcal{D}\,n}(x;\bm{\lambda})$ are
\begin{align}
  &\mathcal{A}_{\mathcal{D}}(\bm{\lambda})
  \phi_{\mathcal{D}\,n}(x;\bm{\lambda})
  =\kappa^{\frac{M}{2}}f_n(\bm{\lambda})
  \phi_{\mathcal{D}\,n-1}(x;\bm{\lambda}+\bm{\delta}),\n
  &\mathcal{A}_{\mathcal{D}}(\bm{\lambda})^{\dagger}
  \phi_{\mathcal{D}\,n-1}(x;\bm{\lambda}+\bm{\delta})
  =\kappa^{-\frac{M}{2}}b_{n-1}(\bm{\lambda})
  \phi_{\mathcal{D}\,n}(x;\bm{\lambda}).
  \label{ADphiDn=,ADdphiDn=}
\end{align}
The forward and backward shift operators are defined by
\begin{align}
  \mathcal{F}_{\mathcal{D}}(\bm{\lambda})&\eqdef
  \psi_{\mathcal{D}}\,(x;\bm{\lambda}+\bm{\delta})^{-1}\circ
  \mathcal{A}_{\mathcal{D}}(\bm{\lambda})\circ
  \psi_{\mathcal{D}}\,(x;\bm{\lambda})\n
  &=\frac{i}{\varphi(x)\check{\Xi}_{\mathcal{D}}(x;\bm{\lambda})}
  \Bigl(\check{\Xi}_{\mathcal{D}}(x+i\tfrac{\gamma}{2};
  \bm{\lambda}+\bm{\delta})e^{\frac{\gamma}{2}p}
  -\check{\Xi}_{\mathcal{D}}(x-i\tfrac{\gamma}{2};
  \bm{\lambda}+\bm{\delta})e^{-\frac{\gamma}{2}p}\Bigr),
  \label{calFD}\\
  \mathcal{B}_{\mathcal{D}}(\bm{\lambda})&\eqdef
  \psi_{\mathcal{D}}\,(x;\bm{\lambda})^{-1}\circ
  \mathcal{A}_{\mathcal{D}}(\bm{\lambda})^{\dagger}\circ
  \psi_{\mathcal{D}}\,(x;\bm{\lambda}+\bm{\delta})\n
  &=\frac{-i}{\check{\Xi}_{\mathcal{D}}(x;\bm{\lambda}+\bm{\delta})}
  \Bigl(V(x;\bm{\lambda}^{[M_{\I},M_{\II}]})
  \check{\Xi}_{\mathcal{D}}(x+i\tfrac{\gamma}{2};\bm{\lambda})
  e^{\frac{\gamma}{2}p}\n
  &\qquad\qquad\qquad\quad
  -V^*(x;\bm{\lambda}^{[M_{\I},M_{\II}]})
  \check{\Xi}_{\mathcal{D}}(x-i\tfrac{\gamma}{2};\bm{\lambda})
  e^{-\frac{\gamma}{2}p}\Bigr)\varphi(x),
  \label{calBD}
\end{align}
and their actions on $\check{P}_{\mathcal{D},n}(x;\bm{\lambda})$ are
\begin{equation}
  \mathcal{F}_{\mathcal{D}}(\bm{\lambda})
  \check{P}_{\mathcal{D},n}(x;\bm{\lambda})
  =f_n(\bm{\lambda})
  \check{P}_{\mathcal{D},n-1}(x;\bm{\lambda}+\bm{\delta}),
  \ \ \mathcal{B}_{\mathcal{D}}(\bm{\lambda})
  \check{P}_{\mathcal{D},n-1}(x;\bm{\lambda}+\bm{\delta})
  =b_{n-1}(\bm{\lambda})
  \check{P}_{\mathcal{D},n}(x;\bm{\lambda}).\!
  \label{FDPDn=,BDPDn=}
\end{equation}
The similarity transformed Hamiltonian is square root free:
\begin{align}
  \widetilde{\mathcal{H}}_{\mathcal{D}}(\bm{\lambda})
  &\eqdef\psi_{\mathcal{D}}(x;\bm{\lambda})^{-1}\circ
  \mathcal{H}_{\mathcal{D}}(\bm{\lambda})\circ
  \psi_{\mathcal{D}}(x;\bm{\lambda})
  =\mathcal{B}_{\mathcal{D}}(\bm{\lambda})
  \mathcal{F}_{\mathcal{D}}(\bm{\lambda})\n
  &=V(x;\bm{\lambda}^{[M_{\I},M_{\II}]})\,
  \frac{\check{\Xi}_{\mathcal{D}}(x+i\frac{\gamma}{2};\bm{\lambda})}
  {\check{\Xi}_{\mathcal{D}}(x-i\frac{\gamma}{2};\bm{\lambda})}
  \biggl(e^{\gamma p}
  -\frac{\check{\Xi}_{\mathcal{D}}(x-i\gamma;\bm{\lambda}+\bm{\delta})}
  {\check{\Xi}_{\mathcal{D}}(x;\bm{\lambda}+\bm{\delta})}\biggr)\n
  &\quad+V^*(x;\bm{\lambda}^{[M_{\I},M_{\II}]})\,
  \frac{\check{\Xi}_{\mathcal{D}}(x-i\frac{\gamma}{2};\bm{\lambda})}
  {\check{\Xi}_{\mathcal{D}}(x+i\frac{\gamma}{2};\bm{\lambda})}
  \biggl(e^{-\gamma p}
  -\frac{\check{\Xi}_{\mathcal{D}}(x+i\gamma;\bm{\lambda}+\bm{\delta})}
  {\check{\Xi}_{\mathcal{D}}(x;\bm{\lambda}+\bm{\delta})}\biggr),
\end{align}
and the multi-indexed orthogonal polynomials
$\check{P}_{\mathcal{D},n}(x;\bm{\lambda})$ are its eigenpolynomials:
\begin{equation}
  \widetilde{\mathcal{H}}_{\mathcal{D}}(\bm{\lambda})
  \check{P}_{\mathcal{D},n}(x;\bm{\lambda})=\mathcal{E}_n(\bm{\lambda})
  \check{P}_{\mathcal{D},n}(x;\bm{\lambda}).
  \label{tHPDn=}
\end{equation}
Other intertwining relations are
(see \eqref{xiDl->l+d}--\eqref{xiDl+d->lII})
\begin{align}
  \kappa^{\frac12}
  \hat{\mathcal{A}}_{d_1\ldots d_{s+1}}(\bm{\lambda}+\bm{\delta})
  \mathcal{A}_{d_1\ldots d_s}(\bm{\lambda})
  &=\mathcal{A}_{d_1\ldots d_{s+1}}(\bm{\lambda})
  \hat{\mathcal{A}}_{d_1\ldots d_{s+1}}(\bm{\lambda}),
  \label{hAA=AhA}\\
  \kappa^{-\frac12}
  \hat{\mathcal{A}}_{d_1\ldots d_{s+1}}(\bm{\lambda})
  \mathcal{A}_{d_1\ldots d_s}(\bm{\lambda})^{\dagger}
  &=\mathcal{A}_{d_1\ldots d_{s+1}}(\bm{\lambda})^{\dagger}
  \hat{\mathcal{A}}_{d_1\ldots d_{s+1}}(\bm{\lambda}+\bm{\delta}),
  \label{hAAd=AdhA}
\end{align}
with the potential function $\hat{V}_{d_1\ldots d_{s+1}}$
given in \eqref{Vhd1..ds} (with $s\to s+1$)
\begin{align}
  \hat{V}_{d_1\ldots d_{s+1}}(x;\bm{\lambda})
  &=\frac{\Xi_{d_1\ldots d_s}(x+i\frac{\gamma}{2};\bm{\lambda})}
  {\Xi_{d_1\ldots d_s}(x-i\frac{\gamma}{2};\bm{\lambda})}
  \frac{\Xi_{d_1\ldots d_{s+1}}(x-i\gamma;\bm{\lambda})}
  {\Xi_{d_1\ldots d_{s+1}}(x;\bm{\lambda})}\n
  &\quad\times\left\{
  \begin{array}{ll}
  \alpha^{\I}(\bm{\lambda}^{[s_{\I},s_{\II}]})
  V^{\prime\,\I}(x;\bm{\lambda}^{[s_{\I},s_{\II}]})
  &:\text{$d_{s+1}$ is of type $\I$}\\[2pt]
  \alpha^{\II}(\bm{\lambda}^{[s_{\I},s_{\II}]})
  V^{\prime\,\II}(x;\bm{\lambda}^{[s_{\I},s_{\II}]})
  &:\text{$d_{s+1}$ is of type $\II$}
  \end{array}\right.,
  \label{Vhd1ds+1}
\end{align}
where $s_{\I}$ and $s_{\II}$ are the numbers of the type $\I$ and $\II$ states
in $\{d_1,\ldots,d_s\}$, respectively.

Although we have restricted $d_j\geq 1$, there is no obstruction for
deletion of $d_j=0$. Including the level 0 deletion corresponds to $M-1$
virtual states deletion:
\begin{align}
  \check{P}_{\mathcal{D},n}(x;\bm{\lambda})\Bigm|_{d^{\I}_{M_{\I}}=0}
  &=A\,\check{P}_{\mathcal{D}',n}(x;\bm{\lambda}+\tilde{\bm{\delta}}^{\I}),\n
  &\quad\mathcal{D}'=\{d^{\I}_1-1,\ldots,d^{\I}_{M_{\I}-1}-1,
  d^{\II}_1+1,\ldots,d^{\II}_{M_{\II}}+1\},
  \label{dM1=0}\\
  \check{P}_{\mathcal{D},n}(x;\bm{\lambda})\Bigm|_{d^{\II}_{M_{\II}}=0}
  &=B\,\check{P}_{\mathcal{D}',n}(x;\bm{\lambda}+\tilde{\bm{\delta}}^{\II}),\n
  &\quad\mathcal{D}'=\{d^{\I}_1+1,\ldots,d^{\I}_{M_{\I}}+1,
  d^{\II}_1-1,\ldots,d^{\II}_{M_{\II}-1}-1\},
  \label{d'M2=0}
\end{align}
where the proportionality constants $A$ and $B$ are given in
\eqref{A_dM1=0}--\eqref{B_d'M2=0}.
These can be shown by using \eqref{FP=,BP=}, \eqref{xil->l+d},
\eqref{xil->l+dII} etc.
The denominator polynomial $\Xi_{\mathcal{D}}$ behaves similarly due to
\eqref{PD0=XiD}.
This is why we have restricted $d_j\geq 1$.

The exceptional $X_{\ell}$ Wilson and Askey-Wilson orthogonal polynomials
presented in \cite{os17,os20} correspond to the simplest case $M=1$,
$\mathcal{D}=\{\ell\}$ of type $\I$, $\ell\geq 1$:
\begin{align}
  \check{\xi}_{\ell}(x;\bm{\lambda})
  &=\check{\Xi}_{\{\ell^{\I}\}}
  (x;\bm{\lambda}+\ell\bm{\delta}-\tilde{\bm{\delta}}^{\I}),\n
  \check{P}_{\ell,n}(x;\bm{\lambda})
  &=\check{P}_{\{\ell^{\I}\},n}(x;\bm{\lambda}+\ell\bm{\delta}
  -\tilde{\bm{\delta}}^{\I})
  \times\left\{
  \begin{array}{ll}
  -(a_1+a_2+n)^{-1}&:\text{W}\\[2pt]
  (a_1a_2q^n)^{\frac12}(1-a_1a_2q^n)^{-1}&:\text{AW}
  \end{array}\right..
\end{align}

As observed in some multi-indexed Laguerre and Jacobi polynomials
\cite{os25}, it can happen that two systems with different sets
$\mathcal{D}$ turn out to be equivalent.
Namely the denominator polynomials with different sets $\mathcal{D}$
may be proportional to each other.
For example, the denominator polynomial of $k$ deletions of type $\I$
virtual states,
$\mathcal{D}_1=\{m^{\I},(m+1)^{\I},\ldots,(m+k-1)^{\I}\}$,
and that of $m$ deletions of type $\II$ virtual states,
$\mathcal{D}_2=\{k^{\II},(k+1)^{\II},\ldots,(k+m-1)^{\II}\}$,
are related,
\begin{equation}
  \check{\Xi}_{\mathcal{D}_1}(x;\bm{\lambda}+m\tilde{\bm{\delta}}^{\II})
  =A\,\check{\Xi}_{\mathcal{D}_2}(x;\bm{\lambda}+k\tilde{\bm{\delta}}^{\I})
  \quad(k,m\geq 1),
  \label{Xi_equiv}
\end{equation}
where the proportionality constant $A$ is given in \eqref{A_equiv}.
{}From this and \eqref{VD2}, we have
\begin{equation}
  V_{\mathcal{D}_1}(x;\bm{\lambda}+m\tilde{\bm{\delta}}^{\II})
  =V_{\mathcal{D}_2}(x;\bm{\lambda}+k\tilde{\bm{\delta}}^{\I}).
\end{equation}
Therefore these two systems are equivalent under the shift of parameters.
Classification of the equivalent classes leading to the same polynomials
is a challenging future problem.

For the cases of type $\I$ only ($M_{\I}=M$, $M_{\II}=0$,
$\mathcal{D}=\{d_1,\ldots,d_M\}$), the expressions \eqref{cXiDdef} and
\eqref{cPDndef} are slightly simplified,
\begin{align}
  &\quad\text{W}_{\gamma}[\check{\xi}^{\I}_{d_1},\ldots,
  \check{\xi}^{\I}_{d_M}](x;\bm{\lambda})
  =\varphi_M(x)
  \check{\Xi}_{\mathcal{D}}(x;\bm{\lambda}),\\
  &\quad\nu^{\I}(x;\bm{\lambda}+M\tilde{\bm{\delta}}^{\I})^{-1}
  \text{W}_{\gamma}[\check{\xi}^{\I}_{d_1},\ldots,\check{\xi}^{\I}_{d_M},
  \nu^{\I}\check{P}_n](x;\bm{\lambda})
  =\varphi_{M+1}(x)
  \check{P}_{\mathcal{D},n}(x;\bm{\lambda})\n[2pt]
  &=i^{\frac12M(M+1)}\left|
  \begin{array}{cccc}
  \check{\xi}^{\I}_{d_1}(x^{(M+1)}_1;\bm{\lambda})&\cdots&
  \check{\xi}^{\I}_{d_M}(x^{(M+1)}_1;\bm{\lambda})
  &r^{\I}_1(x^{(M+1)}_1)\check{P}_n(x^{_(M+1)}_1;\bm{\lambda})\\
  \check{\xi}^{\I}_{d_1}(x^{(M+1)}_2;\bm{\lambda})&\cdots&
  \check{\xi}^{\I}_{d_M}(x^{(M+1)}_2;\bm{\lambda})
  &r^{\I}_2(x^{(M+1)}_2)\check{P}_n(x^{(M+1)}_2;\bm{\lambda})\\
  \vdots&\cdots&\vdots&\vdots\\
  \check{\xi}^{\I}_{d_1}(x^{(M+1)}_{M+1};\bm{\lambda})&\cdots&
  \check{\xi}^{\I}_{d_M}(x^{(M+1)}_{M+1};\bm{\lambda})
  &r^{\I}_{M+1}(x^{(M+1)}_{M+1})\check{P}_n(x^{(M+1)}_{M+1};\bm{\lambda})\\
  \end{array}\right|,
\end{align}
where $r^{\I}_j(x)=r^{\I}_j(x;\bm{\lambda},M+1)$.
The cases of type $\II$ only ($M_{\I}=0$, $M_{\II}=M$) are similar.

\subsection{Analyticity and Hermiticity}
\label{sec:analherm}

At the end of this section we comment on the hermiticity of the Hamiltonian
$\mathcal{H}_{\mathcal{D}}$.
The functions $V(x)$, $\phi_0(x)$ and $\check{\mathcal{R}}(x)$ in
\S\,\ref{sec:analy} correspond to $V_{\mathcal{D}}(x;\bm{\lambda})$,
$\phi_{\mathcal{D}\,0}(x;\bm{\lambda})\propto\psi_{\mathcal{D}}(x;\bm{\lambda})
\check{P}_{\mathcal{D},0}(x;\bm{\lambda})$ and
$\check{P}_{\mathcal{D},n}(x;\bm{\lambda})/
\check{P}_{\mathcal{D},0}(x;\bm{\lambda})$,
respectively.
So the function $G(x)$ \eqref{Gdef} becomes (up to an overall constant)
\begin{equation}
  G(x)=\frac{V(x+i\tfrac{\gamma}{2};\bm{\lambda}^{[M_{\I},M_{\II}]})
  \phi_0(x+i\tfrac{\gamma}{2};\bm{\lambda}^{[M_{\I},M_{\II}]})^2}
  {\check{\Xi}_{\mathcal{D}}(x;\bm{\lambda})^2}
  \check{P}_{\mathcal{D},n}(x+i\tfrac{\gamma}{2};\bm{\lambda})
  \check{P}_{\mathcal{D},m}(x-i\tfrac{\gamma}{2};\bm{\lambda}),
\end{equation}
and we have
\begin{align}
  &G(x)-G^*(x)=
  V(x+i\tfrac{\gamma}{2};\bm{\lambda}^{[M_{\I},M_{\II}]})
  \phi_0(x+i\tfrac{\gamma}{2};\bm{\lambda}^{[M_{\I},M_{\II}]})^2
  \times\frac{\mathcal{P}(x)}{\check{\Xi}_{\mathcal{D}}(x;\bm{\lambda})}
  \times\frac{1}{\check{\Xi}_{\mathcal{D}}(x;\bm{\lambda})},\n
  &\mathcal{P}(x)=
  \check{P}_{\mathcal{D},n}(x+i\tfrac{\gamma}{2};\bm{\lambda})
  \check{P}_{\mathcal{D},m}(x-i\tfrac{\gamma}{2};\bm{\lambda})
  -\check{P}_{\mathcal{D},n}(x-i\tfrac{\gamma}{2};\bm{\lambda})
  \check{P}_{\mathcal{D},m}(x+i\tfrac{\gamma}{2};\bm{\lambda}).
\end{align}
{}From \eqref{Vphi0^2} and \eqref{range2}, this $V\phi_0^2$ part
$V(x+i\tfrac{\gamma}{2};\bm{\lambda}^{[M_{\I},M_{\II}]})
\phi_0(x+i\tfrac{\gamma}{2};\bm{\lambda}^{[M_{\I},M_{\II}]})^2$ has
no poles in the rectangular domain $D_{\gamma}$.
The function $\mathcal{P}(x)$ can be divided by
$\check{\Xi}_{\mathcal{D}}(x;\bm{\lambda})$;
$\mathcal{P}(x)/\check{\Xi}_{\mathcal{D}}(x;\bm{\lambda})
=i\varphi(x)\times(\text{polynomial in $\eta(x)$})$.
Thus the potential singularities of $G-G^*$ originate from the zeros of
$\check{\Xi}_{\mathcal{D}}(x;\bm{\lambda})$.
The left hand side of the condition \eqref{cond(iii)'} vanishes
because of $G(x_1+ix)=G^*(x_1-ix)$, $G(x_2+ix)=0=G^*(x_2-ix)$ for W and
$G(x_1+ix)=G^*(x_1-ix)$, $G(x_2+ix)=G^*(x_2-ix)$ for AW,
on the assumption that there is no singularity on the integration paths.
If $\check{\Xi}_{\mathcal{D}}(x;\bm{\lambda})$ has no zeros in $D_{\gamma}$,
the function $G-G^*$ has no poles in $D_{\gamma}$ and
the condition \eqref{cond(iii)'} is satisfied.
If $\check{\Xi}_{\mathcal{D}}(x;\bm{\lambda})$ has zeros in $D_{\gamma}$,
they appear as complex conjugate pairs, $\alpha\pm i\beta$
($x_1\leq\alpha\leq x_2$, $0<\beta\leq\frac12|\gamma|$), because of
$\check{\Xi}^*_{\mathcal{D}}=\check{\Xi}_{\mathcal{D}}$ and
$\check{\Xi}_{\mathcal{D}}(x;\bm{\lambda})\neq 0$ ($x_1\leq x\leq x_2$).
In order to satisfy the condition \eqref{cond(iii)'},
the sum of the residues of $G-G^*$ should vanish.

The term $V_{\mathcal{D}}+V_{\mathcal{D}}^*$ in $\mathcal{H}_{\mathcal{D}}$ is
\begin{align*}
  V_{\mathcal{D}}(x;\bm{\lambda})+V_{\mathcal{D}}^*(x;\bm{\lambda})
  &=V(x;\bm{\lambda}^{[M_{\I},M_{\II}]})\,
  \frac{\check{\Xi}_{\mathcal{D}}(x+i\frac{\gamma}{2};\bm{\lambda})}
  {\check{\Xi}_{\mathcal{D}}(x-i\frac{\gamma}{2};\bm{\lambda})}
  \frac{\check{\Xi}_{\mathcal{D}}(x-i\gamma;\bm{\lambda}+\bm{\delta})}
  {\check{\Xi}_{\mathcal{D}}(x;\bm{\lambda}+\bm{\delta})}\n
  &\quad+V^*(x;\bm{\lambda}^{[M_{\I},M_{\II}]})\,
  \frac{\check{\Xi}_{\mathcal{D}}(x-i\frac{\gamma}{2};\bm{\lambda})}
  {\check{\Xi}_{\mathcal{D}}(x+i\frac{\gamma}{2};\bm{\lambda})}
  \frac{\check{\Xi}_{\mathcal{D}}(x+i\gamma;\bm{\lambda}+\bm{\delta})}
  {\check{\Xi}_{\mathcal{D}}(x;\bm{\lambda}+\bm{\delta})}.
\end{align*}
This does not cause any obstruction for the hermiticity.
The potential singularities in the interval $x_1\leq x\leq x_2$ are
(\romannumeral1) $V(x;\bm{\lambda}^{[M_{\I},M_{\II}]})$ and
$V^*(x;\bm{\lambda}^{[M_{\I},M_{\II}]})$ at $x=x_1,x_2$,
(\romannumeral2) zeros of
$\check{\Xi}_{\mathcal{D}}(x\mp i\frac{\gamma}{2};\bm{\lambda})$
in the denominators,
(\romannumeral3) zeros of
$\check{\Xi}_{\mathcal{D}}(x;\bm{\lambda}+\bm{\delta})$ in the denominators.
For case (\romannumeral1), the singularities cancel out as in the original
case $V(x;\bm{\lambda})+V^*(x;\bm{\lambda})$.
For case (\romannumeral2), we can show
$\check{\Xi}_{\mathcal{D}}(x\mp i\frac{\gamma}{2};\bm{\lambda})\neq 0$
($x_1\leq x\leq x_2$) by using
$\check{\Xi}_{d_1\ldots d_s}(x;\bm{\lambda})\neq 0$,
$\check{\Xi}_{d_1\ldots d_s}(x;\bm{\lambda}+\bm{\delta})\neq 0$
($x_1\leq x\leq x_2$) and \eqref{xiDl->l+d}--\eqref{xiDl+d->lII}.
For case (\romannumeral3), \eqref{V+V*=Vh+Vh*-tE} and \eqref{Vhd1ds+1}
imply that the denominator factor
$\check{\Xi}_{\mathcal{D}}(x;\bm{\lambda}+\bm{\delta})$ disappears, namely
$\check{\Xi}_{\mathcal{D}}(x;\bm{\lambda}+\bm{\delta})$ does not give
singularities.

Thus the Hamiltonian $\mathcal{H}_{\mathcal{D}}$ is well-defined and
hermitian, if $\check{\Xi}_{\mathcal{D}}(x;\bm{\lambda})$ has no zeros
in $D_{\gamma}$ or the residues coming from the zeros of
$\check{\Xi}_{\mathcal{D}}(x;\bm{\lambda})$ cancel. At present we have no
general proof of the cancellation nor generic procedures to restrict the
parameters so that there will be no zeros of
$\check{\Xi}_{\mathcal{D}}(x;\bm{\lambda})$ in the rectangular domain
$D_{\gamma}$.
Existence of such parameter ranges that
$\check{\Xi}_{\mathcal{D}}(x;\bm{\lambda})$ has no zeros in $D_{\gamma}$
can be verified by numerical calculation for small $M$.

\section{Summary and Comments}
\label{summary}
\setcounter{equation}{0}

By following the examples of the multi-indexed Laguerre, Jacobi \cite{os25}
and ($q$-)Racah \cite{os26} polynomials, the multi-indexed Wilson (W) and
Askey-Wilson (AW) polynomials are constructed within the framework of
discrete quantum mechanics with pure imaginary shifts \cite{os13,os24}.
The method is, as in the previous cases, multiple Darboux-Crum transformations
\cite{crum,adler,darb} by using the virtual state solutions.
The virtual state solutions are derived through certain discrete symmetries
of the original Wilson and Askey-Wilson Hamiltonians and by definition they
are not eigenfunctions of the discrete Schr\"odinger equation.
The type $\I$ and $\II$ virtual state solutions are introduced \eqref{xiv=} but
they are not related with specific boundary conditions, in contradistinction
with the multi-indexed Laguerre, Jacobi or ($q$-)Racah cases.
Main emphasis is on the algebraic structure and the difference equations
for the multi-indexed W and AW polynomials, \eqref{FDPDn=,BDPDn=},
\eqref{tHPDn=} etc., which hold for any parameter range.
So far we do not have a comprehensive method to determine the parameter
ranges which ensure the hermiticity of the deformed Hamiltonians and thus
the orthogonality of the multi-indexed W and AW polynomials.
The one-indexed, {\em i.e.\/} $\mathcal{D}=\{\ell\}$, $\ell\ge1$, of type
$\I$ are identical with the exceptional W or AW polynomials reported
earlier \cite{os17,os20}.

\bigskip
Like the other exceptional polynomials, the multi-indexed W and AW
polynomials do not satisfy the three term recurrence relations.
As in the ordinary Sturm-Liouville problems, the multi-indexed orthogonal
polynomial $P_{\mathcal{D},n}(y;\bm{\lambda})$ has $n$ zeros in the
orthogonality range, $0<y<\infty$ (W) or $-1<y<1$ (AW)
(the oscillation theorem).
It is well known that various hypergeometric orthogonal polynomials in
the Askey scheme are obtained from the Wilson and Askey-Wilson polynomials
in certain limits. Similarly, from the multi-indexed W and AW polynomials
presented in the previous section, we can obtain the multi-indexed version
of various orthogonal polynomials, such as the continuous (dual) Hahn, etc.
In that sense, the multi-indexed Wilson polynomials are also obtained from
the multi-indexed Askey-Wilson polynomials.
Here we briefly discuss the limits to the multi-indexed Jacobi and Laguerre
cases. In an appropriate limit the discrete quantum mechanics with pure
imaginary shifts reduces to the ordinary quantum mechanics \cite{os6}.
Explicitly the W and the AW systems reduce to the Laguerre (L) and the
Jacobi (J) systems, respectively in the following way \cite{os6}:
\begin{align}
  \text{W}:&\ \bm{\lambda}=\Bigl(\frac{c^2}{\omega_1},\frac{c^2}{\omega_2},
  g_1,g_2\Bigr),\quad 1=\omega_1+\omega_2,\ \ g=g_1+g_2-\tfrac12,\n
  &\frac{4}{a_1a_2}\mathcal{H}^{\text{W}}\times c^2
  \xrightarrow{c\to\infty}\mathcal{H}^{\text{L}}
  =p^2+x^2+\frac{g(g-1)}{x^2}-1-2g,\\
  \text{AW}:&\ q^{\bm{\lambda}}=(-q^{h_1},-q^{h_2},q^{g_1},q^{g_2}),\quad
  g=g_1+g_2-\tfrac12,\ \ h=h_1+h_2-\tfrac12,\ \ q=e^{-\frac{1}{c}},
  \ \ x=2x^{\text{J}},\n
  &\bigl(a_1a_2a_3a_4q^{-1}\bigr)^{-\frac12}
  \mathcal{H}^{\text{AW}}\times c^2
  \xrightarrow{c\to\infty}\tfrac14\mathcal{H}^{\text{J}},
  \ \ \mathcal{H}^{\text{J}}=(p^J)^2+\frac{g(g-1)}{\sin^2x^J}
  +\frac{h(h-1)}{\cos^2x^J}-(g+h)^2.
\end{align}
The ground state wavefunction $\phi_0(x)$ \eqref{phi0=W,AW} and the
eigenpolynomial $P_n(x)$ \eqref{Pn=W,AW} also reduce to those of L and J
cases after an appropriate overall rescaling.
For the deformed systems we have the same correspondence under the same limit.
For the AW case, the type $\I$ and $\II$ twists \eqref{twist} reduce to those
of the J cases given in \cite{os25}, $(g,h)\to(g,1-h)$ and $(g,h)\to(1-g,h)$.
For the W case, the type $\II$ twist \eqref{twist} reduces to that of the L
case given in \cite{os25}, $g\to 1-g$.
For the type $\I$ of W case, there is subtlety because of negative components
of $\mathfrak{t}^{\I}(\bm{\lambda})=(1-\frac{c^2}{\omega_1},
1-\frac{c^2}{\omega_2},g_1,g_2)$.
For example, in order to obtain the limit of the ground state wavefunction,
we need certain regularization. The limit of type $\I$ becomes unchanging $g$
and effectively changing $x$ to $ix$. 
This corresponds to the type $\I$ of L given in \cite{os25}.
Therefore the deformed W and AW systems reduce to the deformed L and J
systems in \cite{os25}.
The multi-indexed Wilson and Askey-Wilson polynomials \eqref{cPDndef}
reduce to the multi-indexed Laguerre and Jacobi polynomials given in
\cite{os25} after an appropriate overall rescaling.

\bigskip
When all the parameters $a_i$'s are real, we have  four other twists,
\begin{align*}
  &\mathfrak{t}^{(13)}(\bm{\lambda})
  =(1-\lambda_1,\lambda_2,1-\lambda_3,\lambda_4),\quad
  \mathfrak{t}^{(14)}(\bm{\lambda})
  =(1-\lambda_1,\lambda_2,\lambda_3,1-\lambda_4),\\
  &\mathfrak{t}^{(23)}(\bm{\lambda})
  =(\lambda_1,1-\lambda_2,1-\lambda_3,\lambda_4),\quad
  \mathfrak{t}^{(24)}(\bm{\lambda})
  =(\lambda_1,1-\lambda_2,\lambda_3,1-\lambda_4),
\end{align*}
because of the permutation symmetry of the $a_i$'s.
Algebraically, any one of the six twists defines a deformed Hamiltonian.
According to the parameter configuration, {\em e.g.\/}
$a_1<a_2<a_3<a_4$, etc, the compatibility of any two or more twists and
the hermiticity of the resulting multi-indexed Hamiltonians would be
determined. The detailed analysis of these allowed parameter ranges is
beyond the scope of the present paper.

With the present paper, the project of generic construction
of multi-indexed orthogonal polynomials of a single variable is
now complete.
It is a real challenge to pursue the possibility of constructing
multi-indexed orthogonal polynomials of several variables.

\section*{Acknowledgements}

R.\,S. is supported in part by Grant-in-Aid for Scientific Research
from the Ministry of Education, Culture, Sports, Science and Technology
(MEXT), No.23540303 and No.22540186.

\bigskip
\appendix
\renewcommand{\theequation}{\Alph{section}.\arabic{equation}}
\section{Several Formulas}
\label{sec:app}
\setcounter{equation}{0}

In  Appendix we provide several formulas which are not included
in the main text for smooth presentation.

First we give various proportionality constants:
\begin{align}
  &A\text{ in \eqref{PD0=XiD}}=\left\{
  \begin{array}{ll}
  \prod_{j=1}^{M_{\I}}(-a_1-a_2+d^{\I}_j+1)
  \cdot\prod_{j=1}^{M_{\II}}(-a_3-a_4+d^{\II}_j+1)
  &:\text{W}\\[4pt]
  q^{2M_{\I}M_{\II}}
  \prod_{j=1}^{M_{\I}}(a_1a_2q^{-d^{\I}_j-1})^{-\frac12}
  (1-a_1a_2q^{-d^{\I}_j-1})\\[2pt]
  \qquad\ \times
  \prod_{j=1}^{M_{\II}}(a_3a_4q^{-d^{\II}_j-1})^{-\frac12}
  (1-a_3a_4q^{-d^{\II}_j-1})
  &:\text{AW}
  \end{array}\right.,
  \label{A_PD0=XiD}\\[2pt]
  &A\text{ in \eqref{dM1=0}}=\left\{
  \begin{array}{ll}
  (-1)^{M_{\II}+1}(a_1+a_2+n-1)
  \prod_{j=1}^{M_{\I}-1}d^{\I}_j(-a_1-a_2+a_3+a_4+d^{\I}_j+1)
  &:\text{W}\\[4pt]
  (-1)^{M_{\I}-1}(a_1a_2q^{n-1})^{-\frac12}(1-a_1a_2q^{n-1})\\[2pt]
  \quad\times
  \prod_{j=1}^{M_{\I}-1}q^{-\frac12d^{\I}_j}(1-q^{d^{\I}_j})
  (1-a_1^{-1}a_2^{-1}a_3a_4q^{d^{\I}_j+1})\\[2pt]
  \quad\times(a_1^{-1}a_2^{-1}a_3a_4)^{\frac12M_{\II}}
  \prod_{j=1}^{M_{\II}}q^{M_{\I}+j-\frac12d^{\II}_j}
  &:\text{AW}
  \end{array}\right.,
  \label{A_dM1=0}\\
  &B\text{ in \eqref{d'M2=0}}=\left\{
  \begin{array}{ll}
  -(a_3+a_4+n-1)
  \prod_{j=1}^{M_{\II}-1}d^{\II}_j(-a_3-a_4+a_1+a_2+d^{\II}_j+1)
  &:\text{W}\\[4pt]
  (-1)^{M_{\I}+M_{\II}-1}(a_3a_4q^{n-1})^{-\frac12}(1-a_3a_4q^{n-1})\\[2pt]
  \quad\times
  \prod_{j=1}^{M_{\II}-1}q^{-\frac12d^{\II}_j}(1-q^{d^{\II}_j})
  (1-a_3^{-1}a_4^{-1}a_1a_2q^{d^{\II}_j+1})\\[2pt]
  \quad\times(a_1a_2a_3^{-1}a_4^{-1})^{\frac12M_{\I}}
  \prod_{j=1}^{M_{\I}}q^{M_{\II}+j-\frac12d^{\I}_j}
  &:\text{AW}
  \end{array}\right.,
  \label{B_d'M2=0}\\
  &A\text{ in \eqref{Xi_equiv}}=\left\{
  \begin{array}{ll}
  {\displaystyle
  (-1)^{km}
  \frac{\prod_{j=1}^k(-j)^{k-j}}{\prod_{j=1}^m(-j)^{m-j}}
  \frac{\prod_{j=1}^{[\frac12k]}(a_3+a_4-a_1-a_2+2j)_{2k-4j+1}}
  {\prod_{j=1}^{[\frac12m]}(a_1+a_2-a_3-a_4+2j)_{2m-4j+1}}}  
  &:\text{W}\\[16pt]
  {\displaystyle
  (-a_1^{-1}a_2^{-1}a_3a_4)^{km}
  q^{\frac{1}{12}(k-m)(3km-(k-m-1)(k-m+1))}}\\
  {\displaystyle
  \quad\times
  \frac{\prod_{j=1}^k(1-q^j)^{k-j}}{\prod_{j=1}^m(1-q^j)^{m-j}}
  \frac{\prod_{j=1}^{[\frac12k]}(a_1^{-1}a_2^{-1}a_3a_4q^{2j};q)_{2k-4j+1}}
  {\prod_{j=1}^{[\frac12m]}(a_1a_2a_3^{-1}a_4^{-1}q^{2j};q)_{2m-4j+1}}}
  &:\text{AW}
  \end{array}\right..
  \label{A_equiv}
\end{align}

Next we give the coefficients of the highest degree term of the polynomials
$\Xi_{\mathcal{D}}$ and $P_{\mathcal{D},n}$,
\begin{align}
  &\check{\Xi}_{\mathcal{D}}(x;\bm{\lambda})
  =c_{\mathcal{D}}^{\Xi}(\bm{\lambda})\eta(x)^{\ell}
  +(\text{lower order terms}),\n
  &\check{P}_{\mathcal{D}}(x;\bm{\lambda})
  =c_{\mathcal{D},n}^{P}(\bm{\lambda})\eta(x)^{\ell+n}
  +(\text{lower order terms}).
\end{align}
For $\mathcal{D}=\{d^{\I}_1,\ldots,d^{\I}_{M_{\I}},
d^{\II}_1,\ldots,d^{\II}_{M_{\II}}\}$, they are
\begin{align}
  c_{\mathcal{D}}^{\Xi}(\bm{\lambda})&=
  \prod_{j=1}^{M_{\I}}c_{d^{\I}_j}
  \bigl(\mathfrak{t}^{\I}(\bm{\lambda})\bigr)\cdot
  \prod_{j=1}^{M_{\II}}c_{d^{\II}_j}
  \bigl(\mathfrak{t}^{\II}(\bm{\lambda})\bigr)\n
  &\quad\times\left\{
  \begin{array}{ll}
  \prod\limits_{1\leq j<k\leq M_{\I}}(d^{\I}_k-d^{\I}_j)\cdot
  \prod\limits_{1\leq j<k\leq M_{\II}}(d^{\II}_k-d^{\II}_j)\\[10pt]
  \quad\times
  \prod_{j=1}^{M_{\I}}\prod_{k=1}^{M_{\II}}
  (-a_3-a_4-d^{\I}_j+a_1+a_2+d^{\II}_k)
  &:\text{W}\\[6pt]
  \prod\limits_{1\leq j<k\leq M_{\I}}
  \tfrac12q^{\frac12(d^{\I}_j-d^{\I}_k)}(1-q^{d^{\I}_k-d^{\I}_j})\cdot
  \prod\limits_{1\leq j<k\leq M_{\II}}
  \tfrac12q^{\frac12(d^{\II}_j-d^{\II}_k)}(1-q^{d^{\II}_k-d^{\II}_j})\\[12pt]
  \quad\times
  \prod_{j=1}^{M_{\I}}\prod_{k=1}^{M_{\II}}\frac{2}{\sqrt{a_1a_2a_3a_4}}\,
  q^{j+k-2-\frac12(d^{\I}_j+d^{\II}_k)}
  (a_3a_4q^{d^{\I}_j}-a_1a_2q^{d^{\II}_k})
  &:\text{AW}
  \end{array}\right.,
  \label{cXiD}\\
  c_{\mathcal{D},n}^{P}(\bm{\lambda})&=
  c_{\mathcal{D}}^{\Xi}(\bm{\lambda})c_n(\bm{\lambda})\n
  &\quad\times\left\{
  \begin{array}{ll}
  \prod_{j=1}^{M_{\I}}(-a_1-a_2-n+d^{\I}_j+1)\cdot
  \prod_{j=1}^{M_{\II}}(-a_3-a_4-n+d^{\II}_j+1)
  &:\text{W}\\[6pt]
  q^{2M_{\I}M_{\II}}
  \prod_{j=1}^{M_{\I}}(a_1a_2)^{-\frac12}q^{\frac12(d^{\I}_j+1-n)}
  (1-a_1a_2q^{n-d^{\I}_j-1})\\[4pt]
  \quad\times
  \prod_{j=1}^{M_{\II}}(a_3a_4)^{-\frac12}q^{\frac12(d^{\II}_j+1-n)}
  (1-a_3a_4q^{n-d^{\II}_j-1})
  &:\text{AW}
  \end{array}\right..
  \label{cPDn}
\end{align}

The virtual state wavefunction $\tilde{\phi}_{\mathcal{D}\,\text{v}}$
\eqref{phid1..dsn} for $\mathcal{D}=\{d^{\I}_1,\ldots,d^{\I}_{M_{\I}},
d^{\II}_1,\ldots,d^{\II}_{M_{\II}}\}$ is given by
\begin{align}
  \tilde{\phi}_{\mathcal{D}\,\text{v}}(x;\bm{\lambda})
  &=\frac{\check{\Xi}_{\mathcal{D}\,\text{v}}(x;\bm{\lambda})}
  {\sqrt{\check{\Xi}_{\mathcal{D}}(x-i\frac{\gamma}{2};\bm{\lambda})
  \check{\Xi}_{\mathcal{D}}(x+i\frac{\gamma}{2};\bm{\lambda})}}\n
  &\quad\times(a_1a_2)^{\frac12M_{\I}}(a_3a_4)^{\frac12M_{\II}}
  \kappa^{\frac34M(M+1)-\frac12M_{\I}(M_{\I}+1)-\frac12M_{\II}(M_{\II}+1)}\n
  &\quad\times\left\{
  \begin{array}{ll}
  \kappa^{\frac12M_{\II}(M_{\II}-1)-M_{\I}M_{\II}}
  \tilde{\phi}^{\I}_0(x;\bm{\lambda}^{[M_{\I},M_{\II}]})
  &:\text{$\text{v}$ is of type $\I$}\\[2pt]
  \kappa^{\frac12M_{\I}(M_{\I}-1)-M_{\I}M_{\II}}
  \tilde{\phi}^{\II}_0(x;\bm{\lambda}^{[M_{\I},M_{\II}]})
  &:\text{$\text{v}$ is of type $\II$}
  \end{array}\right..
\end{align}

The polynomial $\xi_{\text{v}}$ with parameters $\bm{\lambda}$ and
$\bm{\lambda}+\bm{\delta}$ are related in the following way \cite{os20}:
\begin{align}
  \frac{i}{\varphi(x)}
  \Bigl(v_1^*(x;\bm{\lambda}+\tilde{\bm{\delta}}^{\I})e^{\frac{\gamma}{2}p}
  -v_1(x;\bm{\lambda}+\tilde{\bm{\delta}}^{\I})e^{-\frac{\gamma}{2}p}\Bigr)
  \check{\xi}^{\I}_{\text{v}}(x;\bm{\lambda})
  &=\hat{f}^{\I}_{0,\text{v}}(\bm{\lambda})
  \check{\xi}^{\I}_{\text{v}}(x;\bm{\lambda}+\bm{\delta}),
  \label{xil->l+d}\\
  \frac{-i}{\varphi(x)}
  \Bigl(v_2(x;\bm{\lambda})e^{\frac{\gamma}{2}p}
  -v_2^*(x;\bm{\lambda})e^{-\frac{\gamma}{2}p}\Bigr)
  \check{\xi}^{\I}_{\text{v}}(x;\bm{\lambda}+\bm{\delta})
  &=\hat{b}^{\I}_{0,\text{v}}(\bm{\lambda})
  \check{\xi}^{\I}_{\text{v}}(x;\bm{\lambda}),
  \label{xil+d->l}\\
  \frac{i}{\varphi(x)}
  \Bigl(v_2^*(x;\bm{\lambda}+\tilde{\bm{\delta}}^{\II})e^{\frac{\gamma}{2}p}
  -v_2(x;\bm{\lambda}+\tilde{\bm{\delta}}^{\II})e^{-\frac{\gamma}{2}p}\Bigr)
  \check{\xi}^{\II}_{\text{v}}(x;\bm{\lambda})
  &=\hat{f}^{\II}_{0,\text{v}}(\bm{\lambda})
  \check{\xi}^{\II}_{\text{v}}(x;\bm{\lambda}+\bm{\delta}),
  \label{xil->l+dII}\\
  \frac{-i}{\varphi(x)}
  \Bigl(v_1(x;\bm{\lambda})e^{\frac{\gamma}{2}p}
  -v_1^*(x;\bm{\lambda})e^{-\frac{\gamma}{2}p}\Bigr)
  \check{\xi}^{\II}_{\text{v}}(x;\bm{\lambda}+\bm{\delta})
  &=\hat{b}^{\II}_{0,\text{v}}(\bm{\lambda})
  \check{\xi}^{\II}_{\text{v}}(x;\bm{\lambda}),
  \label{xil+d->lII}
\end{align}
where the functions $v_1$ and $v_2$ are 
\begin{equation}
  v_1(x;\bm{\lambda})\eqdef\left\{
  \begin{array}{ll}
  \!\!\prod_{j=1}^2(a_j+ix)&\!\!:\text{W}\\[2pt]
  \!\!e^{-ix}\prod_{j=1}^2(1-a_je^{ix})&\!\!:\text{AW}
  \end{array}\right.\!\!\!,
  \ \ v_2(x;\bm{\lambda})\eqdef\left\{
  \begin{array}{ll}
  \!\!\prod_{j=3}^4(a_j+ix)&\!\!:\text{W}\\[2pt]
  \!\!e^{-ix}\prod_{j=3}^4(1-a_je^{ix})&\!\!:\text{AW}
  \end{array}\right.\!\!\!,\!\!\!
  \label{v1v2}
\end{equation}
and the constants $\hat{f}_{s,\text{v}}$ and $\hat{b}_{s,\text{v}}$ are
\begin{align}
  &\hat{f}^{\I}_{s,\text{v}}(\bm{\lambda})=\left\{
  \begin{array}{ll}
  a_1+a_2-\text{v}-1&:\text{W}\\
  -q^{\frac12(\text{v}-s)}(1-a_1a_2q^{-\text{v}-1})&:\text{AW}
  \end{array}\right.,
  \ \ \hat{b}^{\I}_{s,\text{v}}(\bm{\lambda})=\left\{
  \begin{array}{ll}
  a_3+a_4+\text{v}&:\text{W}\\
  -q^{-\frac12(\text{v}-s)}(1-a_3a_4q^{\text{v}})&:\text{AW}
  \end{array}\right.,\\
  &\hat{f}^{\II}_{s,\text{v}}(\bm{\lambda})=\left\{
  \begin{array}{ll}
  a_3+a_4-\text{v}-1&:\text{W}\\
  -q^{\frac12(\text{v}-s)}(1-a_3a_4q^{-\text{v}-1})&:\text{AW}
  \end{array}\right.,
  \ \ \hat{b}^{\II}_{s,\text{v}}(\bm{\lambda})=\left\{
  \begin{array}{ll}
  a_1+a_2+\text{v}&:\text{W}\\
  -q^{-\frac12(\text{v}-s)}(1-a_1a_2q^{\text{v}})&:\text{AW}
  \end{array}\right..
\end{align}
These relations are generalized to the denominator polynomials
$\Xi_{d_1\ldots d_s\text{v}}$.
In the rest of Appendix we consider the set $\mathcal{D}=\{d_1,\ldots,d_s\}$,
in which the number of type $\I$ virtual states is $s_{\I}$ and
that of type $\II$ is $s_{\II}$, $s=s_{\I}+s_{\II}$.
When $\text{v}$ is of type $\I$, the denominator polynomials
$\Xi_{\mathcal{D}\,\text{v}}$ with $\bm{\lambda}$ and
$\bm{\lambda}+\bm{\delta}$ are related as
\begin{align}
  &\frac{i}{\varphi(x)\check{\Xi}_{\mathcal{D}}(x;\bm{\lambda})}
  \Bigl(v_1^*(x;\bm{\lambda}^{[s_{\I},s_{\II}]}+\tilde{\bm{\delta}}^{\I})
  \check{\Xi}_{\mathcal{D}}(x+i\tfrac{\gamma}{2};\bm{\lambda}+\bm{\delta})
  e^{\frac{\gamma}{2}p}\n
  &\phantom{\varphi(x)\check{\Xi}_{\mathcal{D}}(x;\bm{\lambda})}
  -v_1(x;\bm{\lambda}^{[s_{\I},s_{\II}]}+\tilde{\bm{\delta}}^{\I})
  \check{\Xi}_{\mathcal{D}}(x-i\tfrac{\gamma}{2};\bm{\lambda}+\bm{\delta})
  e^{-\frac{\gamma}{2}p}\Bigr)
  \check{\Xi}_{\mathcal{D}\,\text{v}}(x;\bm{\lambda})\n
  &\phantom{\varphi(x)\check{\Xi}_{\mathcal{D}}(x;\bm{\lambda})}
  \ =\kappa^{-s_{\II}}\hat{f}^{\I}_{s,\text{v}}(\bm{\lambda})
  \check{\Xi}_{\mathcal{D}\,\text{v}}(x;\bm{\lambda}+\bm{\delta}),
  \label{xiDl->l+d}\\
  &\frac{-i}{\varphi(x)\check{\Xi}_{\mathcal{D}}
  (x;\bm{\lambda}+\bm{\delta})}
  \Bigl(v_2(x;\bm{\lambda}^{[s_{\I},s_{\II}]})
  \check{\Xi}_{\mathcal{D}}(x+i\tfrac{\gamma}{2};\bm{\lambda})
  e^{\frac{\gamma}{2}p}\n
  &\phantom{\varphi(x)\check{\Xi}_{\mathcal{D}}(x;\bm{\lambda}+\bm{\delta})}
  -v_2^*(x;\bm{\lambda}^{[s_{\I},s_{\II}]})
  \check{\Xi}_{\mathcal{D}}(x-i\tfrac{\gamma}{2};\bm{\lambda})
  e^{-\frac{\gamma}{2}p}\Bigr)
  \check{\Xi}_{\mathcal{D}\,\text{v}}(x;\bm{\lambda}+\bm{\delta})\n
  &\phantom{\varphi(x)\check{\Xi}_{\mathcal{D}}(x;\bm{\lambda}+\bm{\delta})}
  \ =\kappa^{s_{\II}}\hat{b}^{\I}_{s,\text{v}}(\bm{\lambda})
  \check{\Xi}_{\mathcal{D}\,\text{v}}(x;\bm{\lambda}).
  \label{xiDl+d->l}
\end{align}
When $\text{v}$ is of type $\II$, they are
\begin{align}
  &\frac{i}{\varphi(x)\check{\Xi}_{\mathcal{D}}(x;\bm{\lambda})}
  \Bigl(v_2^*(x;\bm{\lambda}^{[s_{\I},s_{\II}]}+\tilde{\bm{\delta}}^{\II})
  \check{\Xi}_{\mathcal{D}}(x+i\tfrac{\gamma}{2};\bm{\lambda}+\bm{\delta})
  e^{\frac{\gamma}{2}p}\n
  &\phantom{\varphi(x)\check{\Xi}_{\mathcal{D}}(x;\bm{\lambda})}
  -v_2(x;\bm{\lambda}^{[s_{\I},s_{\II}]}+\tilde{\bm{\delta}}^{\II})
  \check{\Xi}_{\mathcal{D}}(x-i\tfrac{\gamma}{2};\bm{\lambda}+\bm{\delta})
  e^{-\frac{\gamma}{2}p}\Bigr)
  \check{\Xi}_{\mathcal{D}\,\text{v}}(x;\bm{\lambda})\n
  &\phantom{\varphi(x)\check{\Xi}_{\mathcal{D}}(x;\bm{\lambda})}
  \ =\kappa^{-s_{\I}}\hat{f}^{\II}_{s,\text{v}}(\bm{\lambda})
  \check{\Xi}_{\mathcal{D}\,\text{v}}(x;\bm{\lambda}+\bm{\delta}),
  \label{xiDl->l+dII}\\
  &\frac{-i}{\varphi(x)\check{\Xi}_{\mathcal{D}}
  (x;\bm{\lambda}+\bm{\delta})}
  \Bigl(v_1(x;\bm{\lambda}^{[s_{\I},s_{\II}]})
  \check{\Xi}_{\mathcal{D}}(x+i\tfrac{\gamma}{2};\bm{\lambda})
  e^{\frac{\gamma}{2}p}\n
  &\phantom{\varphi(x)\check{\Xi}_{\mathcal{D}}(x;\bm{\lambda}+\bm{\delta})}
  -v_1^*(x;\bm{\lambda}^{[s_{\I},s_{\II}]})
  \check{\Xi}_{\mathcal{D}}(x-i\tfrac{\gamma}{2};\bm{\lambda})
  e^{-\frac{\gamma}{2}p}\Bigr)
  \check{\Xi}_{\mathcal{D}\,\text{v}}(x;\bm{\lambda}+\bm{\delta})\n
  &\phantom{\varphi(x)\check{\Xi}_{\mathcal{D}}(x;\bm{\lambda}+\bm{\delta})}
  \ =\kappa^{s_{\I}}\hat{b}^{\II}_{s,\text{v}}(\bm{\lambda})
  \check{\Xi}_{\mathcal{D}\,\text{v}}(x;\bm{\lambda}).
  \label{xiDl+d->lII}
\end{align}
These relations are used to show \eqref{hAA=AhA}--\eqref{hAAd=AdhA},
and imply the difference equations of $\check{\Xi}_{\mathcal{D}\,\text{v}}$;
when $\text{v}$ is of type $\I$, they are
\begin{align}
  &\biggl(\alpha^{\I}(\bm{\lambda}^{[s_{\I},s_{\II}]})
  V^{\prime\,\I}(x;\bm{\lambda}^{[s_{\I},s_{\II}]})
  \frac{\check{\Xi}_{\mathcal{D}}(x+i\frac{\gamma}{2};\bm{\lambda})}
  {\check{\Xi}_{\mathcal{D}}(x-i\frac{\gamma}{2};\bm{\lambda})}
  e^{\gamma p}\n
  &\quad+\alpha^{\I}(\bm{\lambda}^{[s_{\I},s_{\II}]})
  V^{\prime\,\I*}(x;\bm{\lambda}^{[s_{\I},s_{\II}]})
  \frac{\check{\Xi}_{\mathcal{D}}(x-i\frac{\gamma}{2};\bm{\lambda})}
  {\check{\Xi}_{\mathcal{D}}(x+i\frac{\gamma}{2};\bm{\lambda})}
  e^{-\gamma p}\n
  &\quad-V(x;\bm{\lambda}^{[s_{\I},s_{\II}]})
  \frac{\check{\Xi}_{\mathcal{D}}(x+i\frac{\gamma}{2};\bm{\lambda})}
  {\check{\Xi}_{\mathcal{D}}(x-i\frac{\gamma}{2};\bm{\lambda})}
  \frac{\check{\Xi}_{\mathcal{D}}(x-i\gamma;\bm{\lambda}+\bm{\delta})}
  {\check{\Xi}_{\mathcal{D}}(x;\bm{\lambda}+\bm{\delta})}\n
  &\quad-V^*(x;\bm{\lambda}^{[s_{\I},s_{\II}]})
  \frac{\check{\Xi}_{\mathcal{D}}(x-i\frac{\gamma}{2};\bm{\lambda})}
  {\check{\Xi}_{\mathcal{D}}(x+i\frac{\gamma}{2};\bm{\lambda})}
  \frac{\check{\Xi}_{\mathcal{D}}(x+i\gamma;\bm{\lambda}+\bm{\delta})}
  {\check{\Xi}_{\mathcal{D}}(x;\bm{\lambda}+\bm{\delta})}
  \biggr)
  \Xi_{\mathcal{D}\,\text{v}}(x;\bm{\lambda})\n
  &=\tilde{\mathcal{E}}^{\I}_{\text{v}}(\bm{\lambda})
  \Xi_{\mathcal{D}\,\text{v}}(x;\bm{\lambda}),
\end{align}
and when $\text{v}$ is of type $\II$, they are
\begin{align}
  &\biggl(\alpha^{\II}(\bm{\lambda}^{[s_{\I},s_{\II}]})
  V^{\prime\,\II}(x;\bm{\lambda}^{[s_{\I},s_{\II}]})
  \frac{\check{\Xi}_{\mathcal{D}}(x+i\frac{\gamma}{2};\bm{\lambda})}
  {\check{\Xi}_{\mathcal{D}}(x-i\frac{\gamma}{2};\bm{\lambda})}
  e^{\gamma p}\n
  &\quad+\alpha^{\II}(\bm{\lambda}^{[s_{\I},s_{\II}]})
  V^{\prime\,\II*}(x;\bm{\lambda}^{[s_{\I},s_{\II}]})
  \frac{\check{\Xi}_{\mathcal{D}}(x-i\frac{\gamma}{2};\bm{\lambda})}
  {\check{\Xi}_{\mathcal{D}}(x+i\frac{\gamma}{2};\bm{\lambda})}
  e^{-\gamma p}\n
  &\quad-V(x;\bm{\lambda}^{[s_{\I},s_{\II}]})
  \frac{\check{\Xi}_{\mathcal{D}}(x+i\frac{\gamma}{2};\bm{\lambda})}
  {\check{\Xi}_{\mathcal{D}}(x-i\frac{\gamma}{2};\bm{\lambda})}
  \frac{\check{\Xi}_{\mathcal{D}}(x-i\gamma;\bm{\lambda}+\bm{\delta})}
  {\check{\Xi}_{\mathcal{D}}(x;\bm{\lambda}+\bm{\delta})}\n
  &\quad-V^*(x;\bm{\lambda}^{[s_{\I},s_{\II}]})
  \frac{\check{\Xi}_{\mathcal{D}}(x-i\frac{\gamma}{2};\bm{\lambda})}
  {\check{\Xi}_{\mathcal{D}}(x+i\frac{\gamma}{2};\bm{\lambda})}
  \frac{\check{\Xi}_{\mathcal{D}}(x+i\gamma;\bm{\lambda}+\bm{\delta})}
  {\check{\Xi}_{\mathcal{D}}(x;\bm{\lambda}+\bm{\delta})}
  \biggr)
  \Xi_{\mathcal{D}\,\text{v}}(x;\bm{\lambda})\n
  &=\tilde{\mathcal{E}}^{\II}_{\text{v}}(\bm{\lambda})
  \Xi_{\mathcal{D}\,\text{v}}(x;\bm{\lambda}).
\end{align}



\begin{thebibliography}{99}
%

\bibitem{os25}
S.\,Odake and R.\,Sasaki,
``Exactly Solvable Quantum Mechanics and Infinite Families of
Multi-indexed Orthogonal Polynomials,''
Phys. Lett. {\bf B702} (2011) 164-170,
{\tt arXiv:\hspace{0pt}1105.0508[math-ph]}.

\bibitem{os26}
S.\,Odake and R.\,Sasaki,
``Multi-indexed ($q$-)Racah Polynomials,''
J. Phys. {\bf A45} (2012) 385201 (21pp),
{\tt arXiv:1203.5868[math-ph]}.

\bibitem{os13}
S.\,Odake and R.\,Sasaki,
``Exactly solvable `discrete' quantum mechanics;
shape invariance, Heisenberg solutions,
annihilation-creation operators and coherent states,''
Prog. Theor. Phys. {\bf 119} (2008) 663-700.
{\tt arXiv:0802.1075[quant-ph]}.

\bibitem{askey}
G.\,E.\,Andrews, R.\,Askey and R.\,Roy,
{\it Special Functions\/},
vol. 71 of Encyclopedia of mathematics and its applications,
Cambridge Univ. Press, Cambridge, (1999).

\bibitem{ismail}
M.\,E.\,H.\,Ismail,
{\it Classical and quantum orthogonal polynomials in one variable\/},
vol. 98 of Encyclopedia of mathematics and its applications,
Cambridge Univ. Press, Cambridge, (2005).

\bibitem{koeswart}
R.\,Koekoek and R.\,F.\,Swarttouw,
``The Askey-scheme of hypergeometric orthogonal polynomials and
its $q$-analogue,''
{\tt arXiv:math.CA/9602214}.

\bibitem{gasper}
G.\,Gasper and M.\,Rahman,
{\it Basic Hypergeometric Series} (2nd ed.),
vol. 96 of Encyclopedia of mathematics and its applications,
Cambridge Univ. Press, Cambridge, (2004).

\bibitem{gomez}
D.\,G\'{o}mez-Ullate, N.\,Kamran and R.\,Milson,
``An extension of Bochner's problem: exceptional invariant subspaces,''
J. Approx Theory {\bf 162} (2010) 987-1006,
{\tt arXiv:0805.\hspace{0pt}3376[math-ph]};
``An extended class of orthogonal polynomials defined by a
Sturm-Liouville problem,''
J. Math. Anal. Appl. {\bf 359} (2009) 352-367,
{\tt arXiv:0807.3939\hspace{0pt}[math-ph]}.

\bibitem{quesne}
C.\,Quesne,
``Exceptional orthogonal polynomials, exactly solvable potentials
and supersymmetry,''
J. Phys. {\bf A41} (2008) 392001,
{\tt arXiv:0807.4087[quant-ph]};
B.\,Bagchi, C.\,Quesne and R.\,Roychoudhury,
``Isospectrality of conventional and new extended potentials,
second-order supersymmetry and role of PT symmetry,''
Pramana J. Phys. {\bf 73} (2009) 337-347,
{\tt arXiv:0812.1488[quant-ph]}.

\bibitem{os16}
S.\,Odake and R.\,Sasaki,
``Infinitely many shape invariant potentials and new orthogonal
polynomials,''
Phys. Lett. {\bf B679} (2009) 414-417,
{\tt arXiv:0906.0142[math-ph]}.

\bibitem{quesne2}
C.\,Quesne,
``Solvable rational potentials and exceptional orthogonal polynomials
in supersymmetric quantum mechanics,''
SIGMA {\bf 5} (2009) 084,
{\tt arXiv:0906.2331[math-\hspace{0pt}ph]}.

\bibitem{os17}
S.\,Odake and R.\,Sasaki,
``Infinitely many shape invariant discrete quantum mechanical systems
and new exceptional orthogonal polynomials related to the Wilson and
Askey-Wilson polynomials,''
Phys. Lett. {\bf B682} (2009) 130-136,
{\tt arXiv:0909.3668[math-ph]}.

\bibitem{os18}
S.\,Odake and R.\,Sasaki,
``Infinitely many shape invariant potentials and cubic identities
of the Laguerre and Jacobi polynomials,''
J. Math. Phys. {\bf 51} (2010) 053513 (9pp),
{\tt arXiv:0911.1585[math-ph]}.

\bibitem{os19}
S.\,Odake and R.\,Sasaki,
``Another set of infinitely many exceptional ($X_{\ell}$) Laguerre
polynomials,''
Phys. Lett. {\bf B684} (2010) 173-176,
{\tt arXiv:0911.3442[math-ph]}.
(Remark: J1(J2) in this reference corresponds to J2(J1) in later references.)

\bibitem{midyaroy}
B.\,Midya and B.\,Roy,
``Exceptional orthogonal polynomials and exactly solvable potentials
in position dependent mass Schr\"odinger Hamiltonians,''
Phys. Lett. {\bf A 373} (2009) 4117-4122,
{\tt arXiv:0910.1209[quant-ph]}.

\bibitem{hos}
C.-L.\,Ho, S.\,Odake and R.\,Sasaki,
``Properties of the exceptional ($X_{\ell}$) Laguerre and Jacobi
polynomials,''
SIGMA {\bf 7} (2011) 107 (24pp),
{\tt arXiv:0912.5447[math-ph]}.

\bibitem{gomez2}
D.\,G\'{o}mez-Ullate, N.\,Kamran and R.\,Milson,
``Exceptional orthogonal polynomials and the Darboux transformation,''
J. Phys. {\bf A43} (2010) 434016 (16pp),
{\tt arXiv:1002.2666\hspace{0pt}[math-ph]};
``On orthogonal polynomials spanning a nonstandard flag,''
Contemp. Math. {\bf 563} (2012) 51--70,
{\tt arXiv:1101.\hspace{0pt}5584[math-ph]}.

\bibitem{stz}
R.\,Sasaki, S.\,Tsujimoto and A.\,Zhedanov,
``Exceptional Laguerre and Jacobi polynomials and the corresponding
potentials through Darboux-Crum transformations,''
J. Phys. {\bf A43} (2010) 315204 (20pp),
{\tt arXiv:1004.4711[math-ph]}.

\bibitem{os20}
S.\,Odake and R.\,Sasaki,
``Exceptional Askey-Wilson type polynomials through Darboux-Crum
transformations,''
J. Phys. {\bf A43} (2010) 335201 (18pp),
{\tt arXiv:1004.0544[math-\hspace{0pt}ph]}.

\bibitem{os21}
S.\,Odake and R.\,Sasaki,
``A new family of shape invariantly deformed Darboux-P\"oschl-Teller
potentials with continuous $\ell$,''
J. Phys. {\bf A44} (2011) 195203 (14pp),
{\tt arXiv:1007.\hspace{0pt}3800[math-ph]}.

\bibitem{ho}
C-L.\,Ho,
``Dirac(-Pauli), Fokker-Planck equations and exceptional Laguerre
polynomials,''
Ann. Phys. {\bf 326} (2011) 797-807,
{\tt arXiv:1008.0744[quant-ph]}.

\bibitem{hs4}
C-L.\,Ho and R.\,Sasaki,
``Zeros of the exceptional Laguerre and Jacobi polynomials,''
ISRN Mathematical Physics, Volume 2012, Article ID 920475 (27pp),
{\tt arXiv:1102.5669\hspace{0pt}[math-ph]}.

\bibitem{os23}
S.\,Odake and R.\,Sasaki,
``Exceptional ($X_{\ell}$) ($q$)-Racah polynomials,''
Prog. Theor. Phys. {\bf 125} (2011) 851-870,
{\tt arXiv:1102.0813[math-ph]}.

\bibitem{Grandati}
Y.\,Grandati,
``Solvable rational extensions of the isotonic oscillator,''
Ann. Phys. {\bf 326} (2011) 2074-2090,
{\tt arXiv:1101.0055\hspace{0pt}[math-ph]};
``Solvable rational extensions of the Morse and Kepler-Coulomb potentials,''
{\tt arXiv:1103.5023[math-ph]}.

\bibitem{ho2}
C-L.\,Ho,
``Prepotential approach to solvable rational potentials and exceptional
orthogonal polynomials,''
Prog. Theor. Phys. {\bf 126} (2011) 185-201,
{\tt arXiv:1104.3511[math-ph]}.

\bibitem{os24}
S.\,Odake and R.\,Sasaki,
``Discrete quantum mechanics,''
J. Phys. A: Math. Theor. {\bf 44} (2011) 353001 (47pp),
{\tt arXiv:1104.0473[math-ph]}.

\bibitem{gomez3}
D.\,G\'{o}mez-Ullate, N.\,Kamran and R.\,Milson,
``Two-step Darboux transformations and exceptional Laguerre polynomials,''
J. Math. Anal. Appr. {\bf 387} (2012) 410-418,
{\tt arXiv:\hspace{0pt}1103.5724[math-ph]}.

\bibitem{takemura}
K.\,Takemura,
``Heun's equation, generalized hypergeometric function and exceptional
Jacobi polynomial,''
J. Phys. A: Math. Theor. {\bf 45} (2012) 085211 (14pp), 
{\tt arXiv:1106.\hspace{0pt}1543[math.CA]}.

\bibitem{quesne3}
C. Quesne,
``Higher-order SUSY, exactly solvable potentials, and exceptional
orthogonal polynomials,''
Mod. Phys. Lett. A {\bf 26} (2011) 1843-1852,
{\tt arXiv:1106.1990[math-ph]};
``Rationally-extended radial oscillators and Laguerre exceptional
orthogonal polynomials in kth-order SUSYQM,''
J. Mod. Phys. A {\bf 26} (2011) 5337-5347,
{\tt arXiv:1110.3958\hspace{0pt}[math-ph]};
``Exceptional orthogonal polynomials and new exactly solvable potentials
in quantum mechanics,''
{\tt arXiv:1111.6467[math-ph]}.

\bibitem{os12}
S.\,Odake and R.\,Sasaki,
``Orthogonal Polynomials from Hermitian Matrices,''
J. Math. Phys. {\bf 49} (2008) 053503 (43pp),
{\tt arXiv:0712.4106[math.CA]}.
(The dual $q$-Meixner polynomial in \S\,5.2.4 and dual $q$-Charlier
polynomial in \S\,5.2.8 should be deleted because the hermiticity of
the Hamiltonian is lost for these two cases.)

\bibitem{os14}
S.\,Odake and R.\,Sasaki,
``Unified theory of exactly and quasi-exactly solvable `discrete'
quantum mechanics: I. Formalism,''
J. Math. Phys {\bf 51} (2010) 083502 (24pp),
{\tt arXiv:\hspace{0pt}0903.2604[math-ph]}.

\bibitem{bochner}
E.\,Routh,
``On some properties of certain solutions of a differential equation
of the second order,''
Proc. London Math. Soc. {\bf 16} (1884) 245-261;
S.\,Bochner,
``\"Uber Sturm-Liouvillesche Polynomsysteme,''
Math. Zeit. {\bf 29} (1929) 730-736.

\bibitem{crum}
M.\,M.\,Crum,
``Associated Sturm-Liouville systems,''
Quart. J. Math. Oxford Ser. (2) {\bf 6} (1955) 121-127,
{\tt arXiv:physics/9908019}.

\bibitem{adler}
M.\,G.\,Krein,
``On continuous analogue of a formula of Christoffel from the theory
of orthogonal polynomials,'' 
Doklady Acad. Nauk. CCCP, {\bf 113} (1957) 970-973;
V.\,\'E.\,Adler,
``A modification of Crum's method,''
Theor. Math. Phys. {\bf 101} (1994) 1381-1386.

\bibitem{Nsusy}
A.\,A.\,Andrianov, M.\,V.\,Ioffe and V.\,P.\,Spiridonov,
``Higher-derivative supersymmetry and the Witten index,''
Phys. Lett. {\bf A 174} (1993) 273-279;
%
H.\,Aoyama, M.\,Sato and T.\,Tanaka,
``General forms of a $\mathcal{N}$-fold supersymmetric family,''
Phys. Lett. {\bf B 503} (2001) 423-429,
{\tt arXiv:quant-ph/0012065};
%
D.\,J.\,Fern\'{a}ndez  and C.\,V.\,Hussin,
``Higher-order SUSY, linearized nonlinear Heisenberg algebras and
coherent states,''
J. Phys. {\bf A 32} (1999) 3603-3619;
%
V.\,G.\,Bagrov and B.\,F.\,Samsonov,
``Supersymmetry of a nonstationary Schr\"odinger equation,''
Phys. Lett. {\bf A 210} (1996) 60-64.

\bibitem{darb}
G.\,Darboux,
{\it Th\'eorie g\'en\'erale des surfaces}
vol 2 (1888) Gauthier-Villars, Paris.

\bibitem{os15}
S.\,Odake and R.\,Sasaki,
``Crum's theorem for `discrete' quantum mechanics,''
Prog. Theor. Phys. {\bf 122} (2009) 1067-1079,
{\tt arXiv:0902.2593[math-ph]}.

\bibitem{gos}
L.\,Garc\'ia-Guti\'errez, S.\,Odake and R.\,Sasaki,
``Modification of Crum's Theorem for `Discrete' Quantum Mechanics,''
Prog. Theor. Phys. {\bf 124} (2010) 1-26,
{\tt arXiv:1004.0289\hspace{0pt}[math-ph]}.

\bibitem{os22}
S.\,Odake and R.\,Sasaki,
``Dual Christoffel transformations,''
Prog. Theor. Phys. {\bf 126} (2011) 1-34,
{\tt arXiv:1101.5468[math-ph]}.

\bibitem{grun-haine}
F.\,A.\,Gr\"unbaum, L.\,Haine, E.\,Horozov,
``Some functions that generalize the Krall-Laguerre polynomials,"
J. Comput. Appl. Math. {\bf 106} (1999) 271-297.

\bibitem{grun-yakim}
F.\,A.\,Gr\"unbaum, M.\,Yakimov,
``Discrete bispectral Darboux transformations from Jacobi operators,"
Pacific J. Math. {\bf 204} (2002) 395-431,
{\tt arXiv:math.CA/0012191}.

\bibitem{haine-ill}
L.\,Haine, P.\,Illiev,
``Askey-Wilson type functions with bound states,"
Ramanujan J. {\bf 11} (2006) 285-329,
{\tt arXiv:math.QA/0203136}.

\bibitem{os7}
S.\,Odake and R.\,Sasaki,
``Unified theory of annihilation-creation operators for solvable
(`discrete') quantum mechanics,''
J. Math. Phys. {\bf 47} (2006) 102102 (33pp),
{\tt arXiv:\hspace{0pt}quant-ph/0605215};
``Exact solution in the Heisenberg picture and annihilation-creation
operators,''
Phys. Lett. {\bf B641} (2006) 112-117,
{\tt arXiv:quant-ph/0605221}.

\bibitem{genden}
L.\,E.\,Gendenshtein,
``Derivation of exact spectra of the Schroedinger equation by means of
supersymmetry,''
JETP Lett. {\bf 38} (1983) 356-359.

\bibitem{os6}
S.\,Odake and R.\,Sasaki,
``Calogero-Sutherland-Moser Systems, Ruijsenaars-Schneider-van Diejen
Systems and Orthogonal Polynomials,"
Prog. Theor. Phys. {\bf 114} (2005) 1245-1260,
{\tt arXiv:hep-th/0512155};
%
``Equilibrium Positions and eigenfunctions of shape invariant
(`discrete') quantum mechanics,"
Rokko Lectures in Mathematics (Kobe University) {\bf 18} (2005) 85-110
(Elliptic Integrable Systems, Eds. M.\,Noumi and K.\,Takasaki),
{\tt arXiv:hep-th/0505070}.

\end{thebibliography}
\end{document}